\documentclass[a4paper,11pt]{article}
\usepackage{jheppub}

\usepackage{amsmath,graphicx,verbatim,epsfig}

\usepackage[utf8x]{inputenc}
\usepackage{lmodern}
\usepackage{amsmath,graphicx}
\usepackage{color}
\usepackage[colorlinks=true,linkcolor=blue,citecolor=blue]{hyperref}

\newcommand{\eps}{\varepsilon}

\newcommand{\nn}{\nonumber}

\newcommand{\bn}{{\bar n}}
\newcommand{\bv}{{\bar v}}
\newcommand{\pslash}{{\not \!p}}
\newcommand{\kslash}{{\not \!k}}
\newcommand{\nslash}{{\not \!n}}

\newcommand{\bnslash}{{\not \!\bn}}

\newcommand{\nb}{\bar n}
\newcommand{\veir}{\varepsilon_{\rm {IR}}}
\newcommand{\veuv}{\varepsilon_{\rm {UV}}}

\newcommand{\be}{\begin{equation}}
\newcommand{\ee}{\end{equation}}
\newcommand{\bea}{\begin{eqnarray}}
\newcommand{\eea}{\end{eqnarray}}
\newcommand{\balign}{\begin{align}}
\newcommand{\ealign}{\end{align}}
\newcommand{\as}{\alpha_s}

\newcommand{\sandwich}[3]{\left< #1 \right | #2 \left | #3 \right >}

\newcommand{\bg}{\begin{gather}}
\newcommand{\foma}{\end{gather}}

\newcommand{\noopsort}[1]{}

\def\e{\epsilon}

\def\ve{\varepsilon}

\def\pd{\partial}

\def\L{\Lambda}

\def\z{\zeta}

\def\S{\Sigma}

\def\<{\langle}
\def\>{\rangle}

\def\a{\alpha}
\def\b{\beta}
\def\g{\gamma}  \def\G{\Gamma}
\def\d{\delta}  \def\D{\Delta}
\def\l{\lambda}   \def\L{\Lambda}
\def\s{\sigma}
  
\def\x{\xi}

\def\m{\mu}
\def\n{\nu}
\def\t{\tau}

\def\z{\zeta}

\def\({\left(}
\def\[{\left[}
\def\){\right)}
\def\]{\right]}

\def\ln{\hbox{ln}}

\def\nslash{n\!\!\!\slash}
\def\bnslash{\bar n\!\!\!\slash}
\def\pslash{p\!\!\!\slash}
\def\bpslash{\bar p\!\!\!\slash}

\def\kslash{k\!\!\!\slash}

\def \le { \left    }
\def \ri { \right }

\def\bp{\bar p}
\def\bP{\bar P}

\title{Factorization Theorem For Drell-Yan At Low $q_T$ And Transverse-Momentum Distributions On-The-Light-Cone}

\author[a]{Miguel G. Echevarr\'ia,}
\author[b]{Ahmad Idilbi}
\author[a]{and Ignazio Scimemi}

\affiliation[a]{Departamento de F\'isica Te\'orica II,
Universidad Complutense de Madrid (UCM), 28040 Madrid, Spain}
\affiliation[b]{Instit\"{u}t f\"{u}r Theoretische Physik, Universit\"at Regensburg, D-93040 Regensburg, Germany}

\emailAdd{miguel.gechevarria@fis.ucm.es}
\emailAdd{ahmad.idilbi@physik.uni-regensburg.de}
\emailAdd{ignazios@fis.ucm.es}

\abstract{
We derive a factorization theorem for Drell-Yan process at low $q_T$ using effective field theory methods. In this theorem all the obtained quantities are gauge invariant and the special role of the soft function--and its subtraction thereof--is emphasized. We define transverse-momentum dependent parton distribution functions (TMDPDFs) which are free from light-cone singularities while all the Wilson lines are defined on-the-light-cone.
 We show explicitly to first order in $\as$ that the partonic Feynman PDF can be obtained from the newly defined partonic TMDPDF by integrating over the transverse momentum of the parton inside the hadron.
We obtain a resummed expression for the TMDPDF, and hence for the cross section, in impact parameter space.
The universality of the newly defined matrix elements is established perturbatively to first order in $\as$.
The factorization theorem is validated to first order in $\alpha_s$ and also the gauge invariance between Feynman and light-cone gauges.

}

\begin{document}
\maketitle

%%%%%%%%%%%%%%%%%%%%%%%%%%%%%%%%%%%%%%%%%
%%%%%%%%%%%%%%%%%%%%%%%%%%%%%%%%%%%%%%%%%
\section{ Introduction}
\label{sec:intro}
%%%%%%%%%%%%%%%%%%%%%%%%%%%%%%%%%%%%%%%%%

Physical observables with non-vanishing (or un-integrated) transverse-momentum dependence are specially important at hadron colliders. Those observables are relevant for the Higgs boson searches and also for proper interpretation of signals of physics ``beyond the Standard Model''. The interest in such observables goes back to the first decade immediately after establishing QCD  as the fundamental theory of strong interactions~\cite{Parisi:1979se,Curci:1979bg,Dokshitzer:1978hw,Collins:1981uk,Collins:1981uw}. Recently however there has been a much renewed interest in $q_T$-differential cross sections where hadrons are involved either in the initial states or in the final ones or in both~\cite{Mantry:2009qz,Mantry:2010bi,Becher:2010tm,Becher:2011xn,Sun:2011iw}. The main issues of interest range from obtaining an appropriate factorization theorem and resumming large logarithmic corrections on one hand and to perform phenomenological predictions both for $q {\bar q}$ and $gg$ production channels on the other.

In this effort we re-examine the derivation of the factorization theorem for Drell-Yan (DY) heavy lepton pair production at small transverse momentum $q_T$ and the proper definition of the non-perturbative matrix elements that arise in such factorization theorems. The region of interest is $\Lambda_{\rm QCD}\ll q_T \ll M$, where $M$ is the heavy lepton pair invariant mass. This topic was considered long time ago by Collins, Soper and Sterman \cite{Collins:1985} where the two notions of factorization and resummation of large logarithms (of the form $\alpha_s^n \ln^m(q_T^2/M^2))$ where systematically investigated. Those efforts yielded the well-known ``CSS formalism''. Here however we implement the techniques of soft-collinear effective theory (SCET) \cite{Bauer:2000yr,Bauer:2001yt,Beneke:2002ph,Bauer:2001ct}
to formally derive the factorization theorem for the $q_T$-differential cross section. Within the framework of effective field theory, other efforts for $q_T$-dependent observables were also considered in \cite{Idilbi:2005er,Mantry:2009qz,Mantry:2010bi,Becher:2010tm,Becher:2011xn}.

Formal manipulations in SCET give us a factorized cross section for Drell-Yan at low $q_T$, which pictorially speaking, looks as follows
\footnote{The leptonic contribution is well-known and it is not shown in this section to simplify the notation.}
\begin{equation}
\label{main}
d\Sigma= H(Q^2/\m_F^2) F_n(\mu_F)\otimes F_{\bn}(\mu_F)\otimes \Phi(\m_F)\,,
\end{equation}
where $Q^2\equiv M^2$ and $\mu_F$ is a factorization scale. Here $H$, $F_{n(\bn)}$ and $\Phi$ stand (respectively) for the hard part, the would-be (two) transverse-momentum dependent parton distribution functions (TMDPDFs) for the two collinear ($n$ and $\bn$) directions of the incoming hadrons and the soft part.
The above result might look familiar and in fact it resembles the one obtained by Ji, Ma and Yuan \cite{Ji:2004wu} for semi-inclusive deep-inelastic scattering (SIDIS) (with the relevant adjustments that need to be made when considering DY instead of SIDIS or vice versa).
In Eq.~(\ref{main}) power corrections of the form $(q_T^2/Q^2)^m$ have been omitted.

Explicit operator definitions for the various quantities in Eq.~(\ref{main}) will be given in the next Sections. However it is worthwhile at this stage to emphasize the important features implied in the derived factorization theorem, Eq.~(\ref {main}). The soft function $\Phi(q_T/Q)$ encodes the effects of emission of soft gluons into the final states with momenta that scale as $Q(\lambda,\lambda,\lambda)$ where $\lambda$ is a small parameter of order $q_T/Q$. Those final state gluons (which hadronize with probability $1$) are needed to kinematically balance the transverse momentum of the produced lepton pair. As we argue below, this function depends only on the transverse coordinates $x_\perp$  and the renormalization scale $\mu$ (which is implicitly assumed). This feature of the soft function is consistent with the definition of the soft functions of Ji. et al. and Collins \cite{Ji:2004wu,Collins:2011ca}. The importance of such soft gluons  was also acknowledged in \cite{Becher:2010tm} however due to the use of a special regulator, the ``analytic regulator'', their contribution vanishes in perturbation theory due to scaleless integrals. It is worth mentioning that in different regulators this is not the case and the soft contribution has to be included--explicitly--in the factorization theorem thus one obtains a regulator-independent theorem as the case should be.

In the effective theory approach soft and collinear partons (with scaling of $Q(1,\lambda^2,\lambda)$ for $n$-collinear or $Q(\lambda^2,1,\lambda)$ for $\bn$-collinear) are not allowed to interact simply because the collinear partons will be driven far off-shell. This is in contrast to ultra-soft and collinear interactions where ultra-soft gluons scale as $Q(\lambda^2,\lambda^2,\lambda^2)$. However ultra-soft gluons are not relevant to the kinematical region of interest \cite{Mantry:2009qz,Becher:2010tm} and will not be discussed further. The relevant framework to describe soft gluons (with $\lambda\sim q_T/Q$) interacting with collinear partons (with off-shellness $Q^2\lambda^2 \gg \Lambda^2_{\rm QCD}$) was named ``SCET-$q_T$'' in \cite{Mantry:2009qz} and here we will adopt the same terminology.
In SCET-$q_T$ the virtuality of the particles is of order $q_T^2$, so it is different from SCET-II~\cite{Bauer:2002aj, Manohar:2006nz}, where the virtuality is of order $\L_{QCD}^2$. SCET-II is needed in order to perform an operator product expansion (OPE) at the intermediate scale $q_T$ which would result in the appearance of the fully integrated PDFs.
In both of these theories soft partons are decoupled from the collinear ones and their mere effect is manifested through the appearance of soft Wilson lines at the level of the matrix elements or Green's functions of the theory.

Due to the fact that the soft function has a non-vanishing contribution in Eq.~(\ref{main}), then one needs immediately to consider the issue of double counting arising from the overlapping regions of soft and collinear modes (when perturbative calculations are performed for the partonic versions of the hadronic matrix elements.) It turns out that this issue will dramatically affect the \emph{proper} definition of the collinear matrix element(s), namely the TMDPDFs. In the traditional perturbative QCD framework the issue of double counting was treated through the notion of ``soft subtraction'' \cite{Collins:2000gd,Collins:1999dz}. In SCET, the analogous treatment was handled through ``zero-bin subtractions'' \cite{Manohar:2006nz}. For sufficiently inclusive observables (and at partonic threshold) an equivalence of the two notions was considered in \cite{Lee:2006nr,Idilbi:2007ff,Idilbi:2007yi}. Also in \cite{Mantry:2009qz,Chiu:2009yx} such equivalence was demonstrated up to ${\cal O}(\alpha_s)$. In our case, we show in Sec.~\ref{sec:zerobin} that for certain IR regulators and in the kinematic region of interest, the equivalence of soft and zero-bin can also be established explicitly to first order in $\alpha_s$.

Given the above and in order to cancel the overlapping contributions, the factorization theorem now reads
\begin{equation}
\label{main2}
d \sigma=H(Q^2/\m^2) [\hat f_n \otimes \phi^{-1}]  \otimes [\hat f_{\bn}\otimes \phi^{-1}] \otimes  \phi\,,
\end{equation}
where the small letter notation refers to the partonic versions of the collinear matrix elements and to the partonic vacuum in the soft function (compared to the QCD one).
The hatted symbols refer to the perturbative calculation of the collinear matrix elements that still include the contamination from the soft momentum region. Variations of the last result appeared in \cite{Ji:2004wu,Chen:2006vd} and also very recently in \cite{Collins:2011ca,Collins:2011zzd} (see also \cite{Aybat:2011zv},\cite{Sun:2011iw}). In SCET one could also consult \cite{Idilbi:2007ff,Chiu:2009yx,Mantry:2009qz,Chay:2005rz,Bauer:2010cc}.

Interestingly enough the last version of the factorization theorem, Eq.~(\ref{main2}), is still problematic. Individually, the partonic collinear matrix elements and the soft function are plagued with un-regularized and un-canceled divergences which render them ill-defined. Those divergences show up perturbatively through integrals of the form:
\begin{equation}
\int_0^1 dt \frac{1}{t}\ .
\end{equation}
which are manifestations of the so-called ``light-cone singularities''. Those divergences appear, for certain IR regulators, also in the (standard) integrated PDFs, however they cancel when combining real and virtual contributions. This is not the case though for TMDPDF. Those light-cone divergences are a result of the fact that the Wilson lines (both soft and collinear) are defined along  light-like trajectories thus allowing for gluons with infinite rapidities to be interacted with. To avoid such singularities, an old idea, due to Collins and Soper, is to tilt the Wilson lines thereby going off-the-light-cone. This trick was pursued in \cite{Ji:2004wu,Chen:2006vd} . More recently, Collins~\cite{Collins:2011zzd} argued that such regulator is necessary to separate ultraviolet (UV) and IR modes thus establishing two purposes: obtaining well-defined objects (free from un-regularized divergences) and a complete factorization of momentum modes.
In the light of the above arguments, one needs to define a new set of collinear matrix elements as follows
\begin{equation}
\label{main3}
j_{n_t(\bn_t)}=\frac{\hat f_{n_t(\bn_t)}}{\sqrt{\phi_t}}\,\,,
\end{equation}
where the subscript $t$ stands for ``tilted'' Wilson lines which are no longer light-like. This is true for collinear and soft ones as well. With this, the factorization theorem takes the form
\begin{equation}
\label{main4}
d\sigma=H(Q^2/\m^2)  j_{n_t} \otimes  j_{\bn_t}\,\,.
\end{equation}

In this work we take a different path. We show that all IR divergences, namely the soft and collinear ones appearing in a massless gauge theory, as well as the light-cone singularities can still be regularized while keeping all the Wilson lines defined on-the-light-cone.
When going off-the-light-cone one introduces the $\zeta$-parameter: $\zeta=(P n_t)^2/(n_t)^2$ where $P$ stands for the incoming hadron momentum. This parameter complicates the phenomenological studies since, among other things, it will affect the evolution of the hadronic matrix elements. However when staying on-the-light-cone, the evolution of the TMDPDF will be governed only by the factorization scale $\m$. Second--and on the technical side--the non-vanishing small components of $n_t$ and ${\bn}_t$ introduce small contributions (in powers of effective theory parameter $\lambda$) that violate the power-counting of that theory unless some {\it{ad-hoc}} relations are imposed between the small and large components of the tilted vectors. It is also not so clear how one can relate the TMDPDF with the integrated PDF when going off-the-light-cone. Moreover, staying on the light-cone is much more compelling when one considers computing, say, the TMDPDF and its anomalous dimension in light-cone gauge. When choosing this gauge then going off-the-light-cone is completely awkward. Those considerations motivate us to stay on-the-light-cone. When doing so one gets
\begin{equation}\label{eq:hjj}
d\sigma=H(Q^2/\m^2)  j_{n}  \otimes j_{\bn}\,\,.
\end{equation}

The above result is still an intermediate step towards getting the final factorization theorem. However, we will \emph{define} our TMDPDFs based on it, where $j_{n(\bn)}$ will be the partonic TMDPDFs, from which one can easily get the hadronic ones.
An extended discussion of the ``on-the-light-cone TMDPDF'' and its properties will be given in the Sections below and it forms the basic contribution of this effort.

 Given that the TMDPDFs, $j_{n(\bn)}$, include soft contribution then they become dependent on the perturbative intermediate scale $q_T$, thus a further step of factorization is needed. This is achieved by performing an operator product expansion (OPE) in impact parameter space in the region $b\ll \L_{QCD}^{-1}$, where the TMDPDFs are matched onto the PDFs ${\cal Q}_{n(\bn)}$.
Once performing the OPE in impact parameter space we get
\footnote{Notice that the convolution in Eq.~(\ref{eq:hjj}) is in momentum space with respect to $\vec k_{n\perp}$ and $\vec k_{\bn\perp}$ while the convolution in Eq.~(\ref{eq:finalfact0}) is in the Bjorken variables $x$ and $z$.}
\begin{align}\label{eq:finalfact0}
d\Sigma= H(Q^2/\m^2)\, \left[\tilde C_n(x;b,Q,\m) \otimes {\cal Q}_n(x;\m)\right]\, \left[\tilde C_\bn(z;b,Q,\m) \otimes {\cal Q}_\bn(z;\m)\right]\,.
\end{align}
Notice that $\tilde C_{n(\bn)}$ still have an explicit $Q^2$-dependence. This dependence is harmless in the sense of factorization of Lorentz invariant scales, since $H$ and $\tilde C_{n(\bn)}$ are both perturbative while ${\cal Q}_{n(\bn)}$ are non-perturbative. However this dependence asks for resummation of logarithms of $Q^2/\m^2$ once $\m$ is chosen to be much smaller that $Q$. The extraction of $Q^2$-dependence of $\tilde C_{n(\bn)}$ and its resummation thereof is discussed in Sec.~\ref{sec:q2} and the final result for the cross section is
\begin{align}\label{eq:finalfact}
d\Sigma = H(Q^2/\m^2)\, &\left[
\left( \frac{Q^2 b^2}{4e^{-2\g_E}} \right)^{-D(\as,L_T)}\tilde {\cal C}_n(x;\vec b_\perp,\m)
\otimes {\cal Q}_n(x;\m)\right]\,
\nn\\
\times
&\left[\left( \frac{Q^2 b^2}{4e^{-2\g_E}} \right)^{-D(\as,L_T)} \tilde {\cal C}_\bn(z;\vec b_\perp,\m)
\otimes {\cal Q}_\bn(z;\m)\right]\,.
\end{align}
This result allows the resummation also for large logarithms of $\L_{QCD}b$ to be performed by simple running between different scales.

Another novel feature of our derived factorization theorem is gauge invariance. Recently it was shown \cite{Idilbi:2010im,GarciaEchevarria:2011md} that SCET, as was traditionally formulated, has to be adjusted by the inclusion of transverse Wilson lines, $T$s, so as to render the basic building blocks and the Lagrangian of SCET gauge invariant under regular and singular gauges. This has the powerful result that all the derived physical quantities (appearing for example in the factorization theorem Eq.~(\ref{main}) or the likes) are gauge invariant and no transverse gauge links need to be invoked by hand in the aftermath. This derivation allows one to consider, for example, the subtracted TMDPDF in covariant gauge, say Feynman gauge, and a singular gauge, say light-cone gauge. We have carried out such computations and found, as expected, full agreement to hold at first order in the strong coupling $\alpha_s$.

It is clear that the soft function $\Phi$ (or $\phi$) connects two different collinear sectors. This might spoil the single-collinearity notion inherited in the standard definitions of the PDFs or TMDPDFs thus the universal features of such quantities might be jeopardized. We discuss this issue extensively and establish the universality of the TMDPDF given in Eq.~(\ref{main3}), in agreement with \cite{Collins:2011zzd}. We also establish, for the first time--and to first order in $\alpha_s$--that when integrating over transverse momentum, the integrated PDF can be recovered from the TMDPDF.

This paper is organized as follows.
In Section~\ref{sec:fac} we establish the factorization theorem for DY at small $q_T$ and the proper definition of the TMDPDF on-the-light-cone.
In Section~\ref{sec:TMDPDF} we calculate $j_{n(\bn)}$ and its anomalous dimension to first order in $\alpha_s$, and discuss the absence of rapidity logarithms in the TMDPDF.
In Section~\ref{sec:tmdpdftopdf} we show explicitly how to obtain the standard PDF from the TMDPDF to first order in $\as$ by integrating over the transverse momentum of the parton.
In Section~\ref{sec:q2} we discuss the $Q^2$-dependence of the intermediate matching coefficients and thus establish the complete factorization of scales. We also show the resummed TMDPDF (and hence the cross section), give its anomalous dimension at second order in $\as$. In Section \ref{sec:universality} we demonstrate that the TMDPDF attains the same order $\alpha_s$ result both in DY and DIS kinematics.
In Section~\ref{sec:factorization} we establish the validity of the factorization theorem to first order in $\alpha_s$.
The gauge invariance of our result is discussed in Section~\ref{sec:lcg} and Section~\ref{sec:zerobin} shows the equivalence of zero-bin and soft function subtraction.
Then we conclude in Section~\ref{sec:conc}.
In Appendix~\ref{sec:app} we show the calculation of the matching of the TMDPDF onto the PDF with $\d$-regulator for DIS and DY kinematics. In Appendix~\ref{sec:app2} we present the calculation of the Quark Form Factor in full QCD with $\d$-regulator for DIS and DY kinematics.

%%%%%%%%%%%%%%%%%%%%%%%%%%%%%%%%%%%%%%%%%
%%%%%%%%%%%%%%%%%%%%%%%%%%%%%%%%%%%%%%%%%
\section{The Factorization of Drell-Yan at Small $q_T$}
\label{sec:fac}
%%%%%%%%%%%%%%%%%%%%%%%%%%%%%%%%%%%%%%%%%
Let the momenta of the two incoming partons initiating the hard reaction be $p$ and $\bp$. A general vector $v$ is decomposed as: $v^\m= \nb \cdot v\frac{n^\m}{2} + n \cdot v\frac{\bn^\m}{2} + v_\perp
=v^+\frac{n^\m}{2} + v^-\frac{\bn^\m}{2} + v_\perp$, where $n=(1,0,0,1)$, $\bn=(1,0,0,-1)$.
\footnote{
Notice that our convention for the light-cone coordinates $p^\pm$ is opposite to the one used in the standard SCET literature.}
We denote $v \equiv |\vec v_\perp|$, and particularly $q_T \equiv |\vec q_\perp|$.
The momentum scalings of the $n$-collinear and $\bn$-collinear were given in the previous Section.

Together these modes give the momentum scaling  of the outgoing photon: $ Q (1,1,\lambda)$.

The initial form of the cross section is
\begin{align}
\label{eq:0}
 d\Sigma &=\frac{4\pi\alpha}{3 q^2 s}\frac{d^4 q}{(2\pi)^4}
 \frac{1}{4}\sum_{\s_1,\s_2} \int d^4 y e^{-i q \cdot y} (-g_{\mu\nu})
\langle N_1(P,\s_1) N_2(\bar P,\s_2)|J^{\mu\dagger}(y)J^\nu(0)|N_1(P,\s_1) N_2(\bar P,\s_2)\rangle ,\nn \\
J^\mu&=\sum_q e_q \bar \psi \g^\m \psi\,,
\end{align}
where $J$ is the electromagnetic current and $e_q$ is the quark electric charge. $P$ and ${\bar P}$ correspond to the hadrons momenta and $s\equiv(P+\bP)^2$. Note that the scaling of the position variable $y$ in Eq.~(\ref{eq:0}) is $y\simeq 1/Q (1,1,1/\lambda)$ and we will make use of this below.

The full QCD current is then matched onto the SCET-$q_T$ one
\begin{align}
\label{eq:jmu}
J^\mu=C(Q^2/\mu^2)\sum_q e_q \bar \chi_\bn S_\bn^{T\dagger} \g^\m S_n^T \chi_n\,,
\end{align}
where in SCET the $n$-collinear and $\nb$-collinear (or anticollinear) fields are described  by
 $\chi_{n(\bn)}=W^{T\dagger}_{n(\bn)}\xi_{n(\bn)}$. For DY kinematics we have
\begin{align}
&W_{n(\bn)}^T =T_{n(\bn)} W_{n(\bn)}\,,
\nn\\
&W_{n}( x) = \bar P \exp \left[i g \int_{-\infty}^0 ds\, \nb \cdot A_n (x+s \bn)\right] \,,
%&\quad&
%W_n^\dagger (x) = P \exp \left[i g \int_{-\infty}^0 ds\, \nb \cdot A_n (x+s n)\right]\,,
%\nn\\
%&W_{\bn} (x) = \bar P \exp \left[-i g \int_{-\infty}^0 ds\, n \cdot A_\bn (x+s \bn)\right] \,,
%&\quad&
%W_{\bn}^\dagger (x) = P \exp \left[i g \int_{-\infty}^0 ds\, n \cdot A_\bn (x+s \bn)\right] \,,
\nn \\
&T_{n} (x) = \bar P \exp \left[i g \int_{-\infty}^0 d\tau\, \vec l_\perp \cdot \vec A_{n\perp} (x^+,\infty^-,\vec x_\perp+\vec l_\perp \tau)\right] \,,
%&\quad&
%T_{n}^\dagger (x) = P \exp \left[i g \int_{-\infty}^0 d\tau l_\perp \cdot A_{n\perp} (x^+,\infty^-,x_\perp+l_\perp \tau)\right] \,,
\nn\\
&T_{\bn} (x) =  \bar P \exp \left[i g \int_{-\infty}^0 d\tau\, \vec l_\perp \cdot \vec A_{\bn\perp} (\infty^+,x^-,\vec x_\perp+\vec l_\perp \tau)\right]\,.
%&\quad&
%T_{\bn}^\dagger (x) =  P \exp \left[i g \int_{-\infty}^0 d\tau \, l_\perp \cdot A_{\bn\perp} (\infty^+,x^-,x_\perp+l_\perp \tau)\right]\,,
\end{align}
$W_{\bn}$ can be obtained from $W_n$ by $n \leftrightarrow \bn$ and $P\leftrightarrow \bP$, where $P$($\bar P$) stands for path (anti-path) ordering.
The transverse Wilson lines are essential to insure gauge invariance of $\chi_{n(\bn)}$ among regular and singular gauges ~\cite{GarciaEchevarria:2011md}.

The soft Wilson lines and their associated transverse Wilson lines are given by
\begin{align}
&S_{n(\bn)}^T = T_{sn(s\bn)} S_{n(\bn)}\,,
\nn\\
&S_n (x) = P \exp \left[i g \int_{-\infty}^0 ds\, n \cdot A_s (x+s n)\right]\,,
%&\quad&
%S_n^\dagger (x) = \bar P \exp \left[-i g \int_{-\infty}^0 ds\, n \cdot A_s (x+s n)\right]\,,
%\nn\\
%&S_\bn (x) = P \exp \left[i g \int_{-\infty}^0 ds  \nb \cdot A_s (x+s \bn)\right]\,,
%&\quad&
%S_\bn^\dagger (x) = \bar P \exp \left[-i g \int_{-\infty}^0 ds  \nb \cdot A_s (x+s \bn)\right]\,,
\nn \\
&T_{sn} (x) = P \exp \left[i g \int_{-\infty}^0 d\tau\, \vec l_\perp \cdot \vec A_{s\perp} (\infty^+,0^-,\vec x_\perp+\vec l_\perp \tau)\right]\,,
%&\quad&
%T_{sn}^\dagger(x)  = \bar P \exp \left[-ig \int_{-\infty}^0 d\tau l_\perp \cdot A_{s\perp} (\infty^+,0^-,x_\perp+l_\perp \tau)\right]\,,
\nn \\
&T_{s\bn} (x) = P \exp \left[i g \int_{-\infty}^0 d\tau\, \vec l_\perp \cdot \vec A_{s\perp} (0^+,\infty^-,\vec x_\perp+\vec l_\perp \tau)\right]\,,
%&\quad&
%T_{s\bn}^\dagger (x) = \bar P \exp \left[-ig \int_{-\infty}^0 d\tau l_\perp \cdot A_{s\perp} (0^+,\infty^-,x_\perp+l_\perp \tau)\right],
\end{align}
where $T_{sn(s\bn)}$ appears for the gauge choice $n \cdot A_s=0$ ($\bn \cdot A_s=0$) and $S_{\bn}$ can be obtained from $S_n$ by $n \leftrightarrow \bn$ and $P\leftrightarrow \bP$.
In the SCET literature there are different ways for obtaining the appropriate soft and collinear Wilson lines. However, one can also start from the full QCD vertex diagram and then take the soft or the collinear limit of the virtual gluon loop momentum. The resulting vertices obtained can unambiguously determine the soft and collinear Wilson lines in the effective theory. The above definitions of the Wilson lines are compatible with the QCD soft and collinear limits for time-like (DY) virtualities.
In Sec.~\ref{sec:zerobin} we present the Wilson lines relevant for space-like (DIS) kinematics and their derivation follows the same argument of taking the soft and collinear limits of QCD.

Using Fierz transformations and  averaging over nucleons spins, the hadronic matrix element  in Eq.~(\ref{eq:0}) can be casted in the form
\begin{align}\label{eq:ma}
 -&\langle N_1(P,\s_1) N_2(\bP,\s_2)|J^{\mu\dagger}(y)J_\mu(0)|N_1(P,\s_1) N_2(\bP,\s_2)\rangle\rightarrow
\nn\\
&
|C(Q^2/\mu)|^2 \sum_q e_q^2 \frac{1}{N_c}
\langle N_1(P,\s_1) N_2(\bP,\s_2)|
\left(\bar \chi_\bn (y) \frac{\bnslash}{2} \chi_\bn(0)\right)
\left(\bar \chi_n(y) \frac{\nslash}{2} \chi_n(0)\right)
\nn\\
&\times
{\rm Tr}\left[\bar {\bf T} (S_n^\dagger(y)S_\bn(y)){\bf T} (S_\bn^\dagger(0)S_n(0))\right]
|N_1(P,\s_1) N_2(\bP,\s_2)\rangle .
\end{align}

Since the $n$-collinear, $\bn$-collinear and soft  fields act  on different Hilbert spaces one can disentangle the Hilbert space itself into a direct product of three distinct Hilbert spaces~\cite{Bauer:2000yr,Lee:2006nr}.
The collinear, anticollinear and  soft fields obey different Lagrangians which are opportunely multipole expanded~\cite{Bauer:2000yr},
however the
multipole expansion of these Lagrangians
does not affect the ``$y$''-dependence of the fields in Eq.~(\ref{eq:ma}) (because there are no interactions
among soft and collinear fields).
Due to these arguments, one can then write the cross section as
\begin{equation}\label{step2}
\begin{aligned}
d\S &= \frac{4\pi\alpha^2}{3 N_c q^2 s}\,\frac{d^4q}{(2\pi)^4}
\int\!d^4y\,e^{-iq\cdot y}\,H(Q^2/\mu^2)\,\sum_q e_q^2\,
F_n(y)\,
F_\bn(y)\,
\Phi(y)\,,
\end{aligned}
\end{equation}
where $H(Q^2/\m^2) = |C(Q^2/\m^2)|^2$ and
\begin{align}
F_n(y) &= \frac{1}{2}\sum_{\s_1}
\langle N_1(P,\s_1)|\,\bar\chi_{n}(y)\,\frac{\rlap{/}{\bar n}}{2}\, \chi_{n}(0)\,|N_1(P,\s_1)\rangle \,,
\nn\\
F_\bn (y) &= \frac{1}{2}\sum_{\s_1}
\langle N_2(\bP,\s_2)|\,\bar\chi_{\overline{n}}(0)\,\frac{\rlap{/}{n}}{2}\,
\chi_{\overline{n}}(y)\,|N_2(\bP,\s_2)\rangle \,,
\nn \\
\Phi(y) &=
\sandwich{0}{ {\rm Tr} \;\bar {\bf T} \big[S_n^{T\dagger} S^T_\bn \big](y){\bf T} \big[S^{T\dagger}_\bn S^T_n\big](0)}{0}\,.
\end{align}
We now Taylor expand Eq.~(\ref{step2}) in the physical limit that we are interested in. The photon is hard with momentum $q\sim Q(1,1,\l)$, so the in exponent $e^{-i q y}$ in Eq.~(\ref{step2}) one has  $y\sim \frac{1}{Q}(1,1,1/\l)$ as mentioned before. On  the other hand   the scaling of the derivatives of the $n$-collinear, anticollinear  and soft terms  are clearly the same as their respective momentum scalings.
%\begin{align}
%(\frac{\pd}{\pd y^-} F_n,\frac{\pd}{\pd y^+} F_n,\frac{\pd}{\pd y_\perp} F_n) &\sim (1,\l^2,\l);
%\quad\quad\quad
%(x^-\frac{\pd}{\pd x^-} f_n,x^+\frac{\pd}{\pd x^+} f_n,x_\perp\frac{\pd}{\pd x_\perp} f_n) &\sim (1,\l^2,1)
%\nn\\
%(\frac{\pd}{\pd y^-} F_\bn,\frac{\pd}{\pd y^+} F_\bn,\frac{\pd}{\pd y_\perp} F_\bn) &\sim (\l^2,1,\l);
%\quad\quad\quad
%(x^-\frac{\pd}{\pd x^-} f_\bn,x^+\frac{\pd}{\pd x^+} f_\bn,x_\perp\frac{\pd}{\pd x_\perp} f_\bn) &\sim (\l^2,1,1)
%\nn\\
%(\frac{\pd}{\pd y^-} \Phi_{\rm DY},\frac{\pd}{\pd y^+} \Phi_{\rm DY},\frac{\pd}{\pd y_\perp} \Phi_{\rm DY}) &\sim (\l,\l,\l);
%\quad\quad\quad
%(x^-\frac{\pd}{\pd x^-} \Phi_{\rm DY},x^+\frac{\pd}{\pd x^+} \Phi_{\rm DY},x_\perp\frac{\pd}{\pd x_\perp} \Phi_{\rm DY}) &\sim (\l,\l,1)
%\end{align}
Combining this with the scaling of $y$, the leading term (${\cal O} (1)$) of the cross section reads
\begin{equation}\label{step3}
\begin{aligned}
d\S &= \frac{4\pi\alpha^2}{3 N_c q^2 s}\,\frac{d^4q}{(2\pi)^4}
\int\!d^4y\,e^{-iq\cdot y}\,H(Q^2/\mu^2)\,\sum_q e_q^2\,
\nn\\
&\times
F_n(0^+,y^-,\vec y_\perp)\,
F_\bn(y^+,0^-,\vec y_\perp)\,
\Phi(0^+,0^-,\vec y_\perp)\,+ {\cal O}(\lambda)\,.
\end{aligned}
\end{equation}
It should be immediately noted that if one had considered ultra-soft scaling $Q(\l^2,\l^2,\l^2)$ instead of the soft one in $\Phi$, then after the Taylor expansion $\Phi$ would be exactly $1$ to all orders in perturbation theory.
This is the case that was considered in \cite{Becher:2010tm}. The fact that the soft function, $\Phi$, depends only on the transverse coordinates is of crucial importance.

In the rest of this Section we will consider the leading order contribution to the partonic version of the cross section Eq.~(\ref{step3}), namely
\begin{align}
d\sigma &= \frac{4\pi\alpha^2}{3 N_c q^2 }\,\frac{dx dz d^2\vec q_\perp}{2(2\pi)^4}
H(Q^2/\mu^2)\,\sum_q e_q^2
\nn\\
&\times
\int d^2\vec k_{n\perp} d^2\vec k_{\bn\perp}d^2\vec k_{s\perp}\,
\d^{(2)}(\vec q_\perp-\vec k_{n\perp}-\vec k_{\bn\perp}-\vec k_{s\perp})\,
f_n(x;\vec k_{n\perp})\,
f_\bn(z;\vec k_{\bn\perp})\,
\phi(\vec k_{s\perp})\,.
\end{align}
with
\begin{align}
f_n(x;\vec k_{n\perp}) &=\frac{1}{2}
\int \frac{dr^-d^2\vec r_\perp}{(2\pi)^3} e^{-i(\frac{1}{2}r^-xp^+-\vec r_\perp \cdot \vec k_{n\perp})}
f_n(0^+,r^-,\vec r_\perp)\,,
\nn\\
f_\bn(z;\vec k_{\bn\perp}) &=\frac{1}{2}
\int \frac{dr^+d^2\vec r_\perp}{(2\pi)^3} e^{-i(\frac{1}{2}r^+z\bp^--\vec r_\perp \cdot \vec k_{\bn\perp})}
f_\bn(r^+,0^-,\vec r_\perp)\,,
\nn\\
\phi(\vec k_{s\perp}) &=
\int \frac{d^2\vec r_\perp}{(2\pi)^2} e^{i \vec r_\perp \cdot \vec k_{s\perp}}
\phi(0^+,0^-,\vec r_\perp)\,.
\label{eq:ftmd}
\end{align}

\emph{Discussion of the TMDPDF:}

The problem with the above factorization formula is that in the ${\cal O}(\alpha_s)$ calculation of $f_{n(\bn)}$ and $\phi$ there are still mixed UV and IR divergences which complicate both the renormalization procedure and the non-perturbative interpretation of such quantities. However, when considering the following combinations
\begin{align}\label{eq:jtmdgeneral}
j_n(x;\vec k_{n\perp}) &=\frac{1}{2}
\int \frac{dr^-d^2\vec r_\perp}{(2\pi)^3} e^{-i(\frac{1}{2}r^-xp^+-\vec r_\perp \cdot \vec k_{n\perp})}
f_n(0^+,r^-,\vec r_\perp)\sqrt{\phi(0^+,0^-,\vec r_\perp)}\,,
\nn\\
j_\bn(z;\vec k_{\bn\perp}) &=\frac{1}{2}
\int \frac{dr^+d^2\vec r_\perp}{(2\pi)^3} e^{-i(\frac{1}{2}r^+z\bp^--\vec r_\perp \cdot \vec k_{\bn\perp})}
f_\bn(r^+,0^-,\vec r_\perp)\sqrt{\phi(0^+,0^-,\vec r_\perp)}\,,
\end{align}
it turns out that those quantities are free from such mixed divergences. This is shown explicitly to hold to ${\cal O} (\alpha_s)$ in the next Section. $j_{n(\bn)}$ in the last equation are our definition of the TMDPDFs.

As mentioned in the Introduction, the equivalence of the zero-bin and soft subtractions is a subtle issue, since it depends on the kinematics and the IR regulator that is used. In Sec~\ref{sec:zerobin} we show this equivalence to first order in $\as$, which leads us to the following relation between $f$ and ${\hat f}$,
\begin{align}
f_n(0^+,r^-,\vec r_\perp) &=
\frac{\hat f_n(0^+,r^-,\vec r_\perp)}{\phi(0^+,0^-,\vec r_\perp)}\,,
\quad\quad\quad
f_\bn(r^+,0^-,\vec r_\perp) =
\frac{\hat f_\bn(r^+,0^-,\vec r_\perp)}{\phi(0^+,0^-,\vec r_\perp)}\,,
\end{align}
where
\begin{align}
\hat f_n(0^+,r^-,\vec r_\perp) &= \langle p|\,\le[\bar\x_n W^T_n\ri](0^+,y^-,\vec y_\perp)\,\frac{\bnslash}{2}\,
\le[W_n^{T\dagger} \x_n\ri](0)\,|p\rangle |_\textrm{zb~included}\,,
\nn\\
\hat f_\bn(r^+,0^-,\vec r_\perp) &= \langle \bar p|\,\le[\bar\x_\bn W^T_\bn\ri](0)\,\frac{\nslash}{2}\,
\le[W_\bn^{T\dagger} \x_\bn\ri](y^+,0^-,\vec y_\perp)\,|\bar p\rangle |_\textrm{zb~included}\,,
\nn\\
\phi(0^+,0^-,\vec r_\perp) &= \sandwich{0}{ {\rm Tr} \; \Big[S_n^{T\dagger} S^T_\bn \Big](0^+,0^-,\vec y_\perp)\le[S^{T\dagger}_\bn S^T_n\ri](0)}{0}\,.
\end{align}

Then, we can write the TMDPDFs in the following way,
\begin{align} \label{eq:jtmd}
j_n(x;\vec k_{n\perp}) &=\frac{1}{2}
\int \frac{dr^-d^2\vec r_\perp}{(2\pi)^3} e^{-i(\frac{1}{2}r^-xp^+-\vec r_\perp \cdot \vec k_{n\perp})}
\frac{\hat f_n(0^+,r^-,\vec r_\perp)}{\sqrt{\phi(0^+,0^-,\vec r_\perp)}}\,,
\nn\\
j_\bn(z;\vec k_{\bn\perp}) &=\frac{1}{2}
\int \frac{dr^+d^2\vec r_\perp}{(2\pi)^3} e^{-i(\frac{1}{2}r^+z\bp^--\vec r_\perp \cdot \vec k_{\bn\perp})}
\frac{\hat f_\bn(r^+,0^-,\vec r_\perp)}{\sqrt{\phi(0^+,0^-,\vec r_\perp)}}\,,
\end{align}
where the square root of the soft function is subtracted from the naive collinear matrix element.

Thus it is compelling to re-cast the factorization theorem in the following form
\begin{align}\label{step42}
d\sigma &= \frac{4\pi\alpha^2}{3 N_c q^2 }\,\frac{dx dz d^2\vec q_\perp}{2(2\pi)^4}
H(Q^2/\mu^2)\,\sum_q e_q^2
\nn\\
&\times
\int d^2\vec k_{n\perp} d^2\vec k_{\bn\perp}\,
\d^{(2)}(\vec q_\perp-\vec k_{n\perp}-\vec k_{\bn\perp})
j_n(x;\vec k_{n\perp},\m)\,
j_\bn(z;\vec k_{\bn\perp},\m)\,,
\end{align}
which is formally in agreement with the line of argument of Refs.~\cite{Collins:2011zzd,Collins:2011ca}. The TMDPDFs $j_{n(\bn)}$ are defined in general in Eq.~(\ref{eq:jtmdgeneral}), but in the following we take the result in Eq.~(\ref{eq:jtmd}), which applies for our particular kinematical regime (perturbative $q_T$ and away from threshold) and for the set of IR regulators implemented below.

The last result is still not the final form of the factorization theorem and the TMDPDFs still have to be refactorized. In the effective theory approach this corresponds to a second step matching of SCET-$q_T$ that describes the physics at the intermediate scale $q_T \gg \L_{QCD}$ with SCET-II that captures the non-perturbative physics at the hadronic scale $\L_{QCD}$.

The refactorization of the TMDPDF is essential since the naive collinear $\hat f_{n(\bn)}$ and the soft contribution $\phi$ that enter in the definition of $j_{n(\bn)}$ live at the intermediate scale $q_T$, consistent with their construction in SCET-$q_T$. Since $q_T$ is perturbative, its conjugate coordinate, the impact parameter $b$, is small enough to perform an OPE in the impact parameter space. Moreover in this space the IR structure becomes manifest with the appearance of IR poles in dimensional regularization. Obviously, the first term in the OPE would be just the standard Feynman PDF, and the Wilson coefficient would be the term that sums all the large logs between $\L_{QCD}$ and $q_T$  (see Refs.~\cite{Collins:1981uk,Collins:1981uw,Collins:1985} and more recently using SCET Refs.~\cite{Mantry:2009qz,Becher:2010tm,Stewart:2009yx,Chiu:2012ir}).
Then, given the following OPE (and an analogous for $\bn$)
\begin{align}
\tilde j_n(x;\vec b_\perp,\m) &= \int_x^1 \frac{dx'}{x'} \tilde C_n\le(\frac{x}{x'};\vec b_\perp,\m\ri)\, {\cal Q}_n(x';\m)
+ {\cal O}(b^2\L^2_{QCD})\,,
\end{align}
where
\begin{align}
\tilde j_n(x;\vec b_\perp,\m) = \int d^2\vec k_{n\perp}\, e^{i\vec k_{n\perp} \cdot \vec b_\perp} j_n(x;\vec k_{n\perp},\m)\,,
\end{align}
the factorization theorem takes the form
\begin{align}\label{eq:mainfact}
d\sigma &= \frac{4\pi\alpha^2}{3 N_c q^2 }\,\frac{dx dz d^2\vec q_\perp}{2(2\pi)^4} \sum_q e_q^2
\int \frac{d^2\vec b_\perp}{(2\pi)^2} e^{-i\vec q_\perp \cdot \vec b_\perp}
\int_x^1 \frac{dx'}{x'} \int_z^1 \frac{dz'}{z'}
\nn\\
&\times
H(Q^2/\mu^2)\, \tilde C_n\le(\frac{x}{x'};\vec b_\perp,Q,\m\ri)\, \tilde C_\bn \le(\frac{z}{z'};\vec b_\perp,Q,\m\ri)\,
{\cal Q}_n(x';\m)\, {\cal Q}_\bn (z';\m)\,.
\end{align}

In this effort and for simplicity of presentation we will not consider the contribution coming from a gluon splitting into two quarks.
This contribution is certainly vital for the final result of the DY cross-section. Here however we are mainly interested in studying the TMDPDF of a quark in a quark. Henceforth  we will refer to this quantity simply as the ``TMDPDF'' and it can be easily checked that all the results below are not affected by this omission.

The above result is one of the main results of this paper and it holds to all orders in perturbation theory. It is worthy to notice the separation of scales: the hard matching coefficient lives at scale $Q$, the matching coefficients at the intermediate scale live at $1/b \sim q_T$, and finally the PDFs live at the hadronic scale $\L_{QCD}$.

As we show below in Sec.~\ref{sec:q2}, $\tilde C_{n(\bn)}$ have a subtle $Q^2$-dependence which at first sight might spoil the scale factorization, however this dependence can be extracted and exponentiated thus putting it under control.

%%%%%%%%%%%%%%%%%%%%%%%%%%%%%%%%
%%%%%%%%%%%%%%%%%%%%%%%%%%%%%%%%
\section{The TMDPDF On-The-Light-Cone}
\label{sec:TMDPDF}
%%%%%%%%%%%%%%%%%%%%%%%%%%%%%%%%

In this section we compute the TMDPDF in Eq.~(\ref{eq:jtmd}) to first order in $\as$ while regularizing the UV divergences in dimensional regularization (DR) and in $\overline{\textrm{MS}}$ scheme.
 The IR divergences as well as the light-cone ones will be regularized by the $\d$-regulator introduced below. We also present a calculation of the virtual part of the TMDPDF in pure DR. This is shown in Subsection~\ref{sec:adpuredr}. Notice that individual Feynman diagrams with real gluon emission cannot be obtained in pure DR due to light-cone singularities and an additional regulator is needed. When such diagrams are added up, the logarithmic dependence on this regulator cancels out which means, at least to first order in $\as$, that
real gluon contribution is free from light-cone divergences. More discussion on this is given below.

We write the poles of the fermion propagators with a real and positive parameters $\D^\pm$
\begin{align}\label{fermionsDelta}
\frac{i(\pslash+\kslash)}{(p+k)^2+i0} \longrightarrow
\frac{i(\pslash+\kslash)}{(p+k)^2+i\D^-}\,,
\quad\quad
\frac{i(\bpslash+\kslash)}{(\bp+k)^2+i0} \longrightarrow
\frac{i(\bpslash+\kslash)}{(\bp+k)^2+i\D^+}\,.
\end{align}

The above prescription applies as well to the fermion propagators in SCET. The corresponding pole-shifting for collinear and soft Wilson lines goes as follows
\begin{align}
\frac{1}{k^+\pm i0} \longrightarrow
\frac{1}{k^+\pm i\d^+}\,,
\quad\quad
\frac{1}{k^-\pm i0} \longrightarrow
\frac{1}{k^-\pm i\d^-}\,,
\end{align}
where $\d^\pm$ are related with $\D^\pm$ through the large components of the collinear fields
\begin{align} \label{regul_DeltaDY}
\d^+ = \frac{\D^+}{\bp^-}\,, \quad &\quad \quad \d^- = \frac{\D^-}{p^+} \,.
\end{align}
The $\d$-regulator resembles the one used in~\cite{Chiu:2009yx}.
Notice that the essential feature of the $\d$-regulator is the coherency between the regulators of fermion propagators and the corresponding Wilson lines.

Expanding Eq.~(\ref{eq:jtmd}) to first order in $\as$ one finds
\begin{align}
\label{eq:jnn}
j_n(x;\vec k_{n\perp},Q,\m)
&=\frac{1}{2}
\int \frac{d\x^-d^2\vec \x_\perp}{(2\pi)^3} e^{-i(\frac{1}{2}\x^-xp^+-\vec \x_\perp \cdot \vec k_{n\perp})}
\left[
\hat f_{n0} + \left(\hat f_{n1} - \frac{1}{2} \hat f_{n0} \phi_1 \right)
\right]+{\cal O}(\a_s^2)\,,
\end{align}
where the numerical subscripts denote the order in the $\a_s$ expansion.
The collinear matrix element at tree level is
\begin{align}\label{tree}
\hat f_{n0} &=  \langle p|\overline{\x}_n(\x^-,0^+,\x_\perp)\, \frac{\bnslash}{2} \,
\x_n(0)|p\rangle\
= e^{i\frac{1}{2}p^+\x^-} p^+ \,,
\end{align}
so the expansion of $j_n$ up to order $\a_s$ is
\begin{align} \label{jn}
j_n(x;\vec k_{n\perp},Q,\m) &=
\d(1-x)\d^{(2)}(\vec k_{n\perp})
\nn\\
&+
\left[
\frac{1}{2}\int \frac{d\x^-d^2\vec \x_\perp}{(2\pi)^3} e^{-i(\frac{1}{2}\x^-xp^+-\vec \x_\perp \cdot \vec k_{n\perp})} \hat f_{n1}
 - \frac{1}{2}\d(1-x)
 \int \frac{d^2\vec \x_\perp}{(2\pi)^2} e^{i\vec \x_\perp \cdot \vec k_{n\perp}} \phi_1
 \right]\,.
\end{align}

%%%%%%%%%%%%%%%%%%%%%%%%%%%%%%%%%
\subsection{Virtual Diagrams}
\label{sec:virtual}
%%%%%%%%%%%%%%%%%%%%%%%%%%%%%%%%%
%%%%%%%%%%%%%%%%%%%%%%%%%%%%%FIGURE
\begin{figure}
\begin{center}
\includegraphics[width=0.8\textwidth]{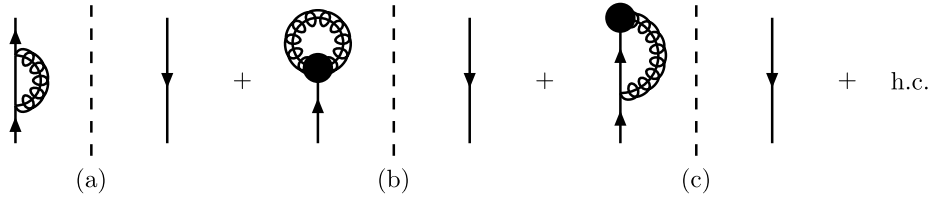}
\end{center}
\caption{\it Virtual corrections for the collinear matrix element. The black blobs represent the collinear
Wilson lines $W$ in Feynman gauge or the $T$ Wilson lines in light-cone gauge. Curly propagators with a line stand for collinear gluons. ``h.c.'' stands for Hermitian conjugate.}
\label{n_virtuals}
\end{figure}
%%%%%%%%%%%%%%%%%%%%%%%%%%%%%%%%%ENDFIGURE
%%%%%%%%%%%%%%%%%%%%%%%%%%%%%%%%%%%%%%%%%%%%FIGURE
\begin{figure}
\begin{center}
\includegraphics[width=0.7\textwidth]{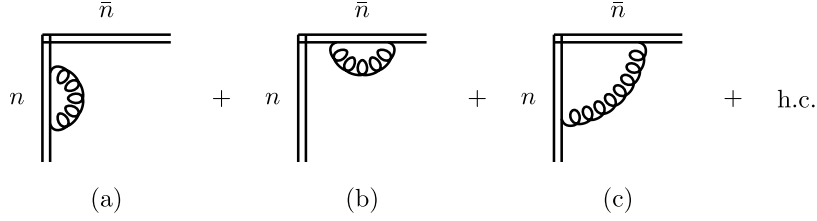}
\end{center}
\caption{\it Virtual corrections for the soft function. Double lines represent the soft  Wilson lines, $S$. ``h.c.'' stands for  Hermitian conjugate.}
\label{s_virtuals}
\end{figure}
%%%%%%%%%%%%%%%%%%%%%%%%%%%%%%%%%ENDFIGURE

The diagrams in figs.~(\ref{n_virtuals}) and~(\ref{s_virtuals}) give collinear  and soft  virtual contributions respectively to $j_n$. The Wave Function Renormalization (WFR) diagram~(\ref{n_virtuals}a) and its Hermitian conjugate give
\begin{align}\label{1a}
\hat f_{n1}^{(\ref{n_virtuals}a)}&=
\frac{\alpha_s C_F}{2\pi}
\d(1-x)\d^{(2)}(\vec k_{n\perp})
\le[ \frac{1}{2\veuv}+\frac{1}{2}\ln\frac{\m^2}{-i\D^-}+\frac{1}{4} \ri] + h.c.
\end{align}
The $W$ Wilson line tadpole diagram, (\ref{n_virtuals}b), is identically $0$, since $\bn^2=0$.
Diagram~(\ref{n_virtuals}c) and its Hermitian conjugate give
\begin{align}
\label{eq:jn1c}
\hat f_{n1}^{(\ref{n_virtuals}c)}&=
-2ig^2C_F \d(1-x)\d^{(2)}(\vec k_{n\perp}) \m^{2\e} \int \frac{d^dk}{(2\pi)^d}
\frac{p^++k^+}{[k^+-i\d^+][(p+k)^2+i\D^-][k^2+i0]}
+ h.c.\nn\\
&=
\frac{\a_s C_F}{2\pi}
\d(1-x)\d^{(2)}(\vec k_{n\perp})
\left[
\frac{2}{\veuv}\ln\frac{\d^+}{p^+} + \frac{2}{\veuv} - \ln^2\frac{\d^+\D^-}{p^+\m^2}
- 2\ln\frac{\D^-}{\m^2} + \ln^2\frac{\D^-}{\m^2} + 2 - \frac{7\pi^2}{12}
\right]\, .
\end{align}
The contribution of diagrams~(\ref{s_virtuals}a) and~(\ref{s_virtuals}b) is zero, since~(\ref{s_virtuals}a) is proportional to $n^2=0$
and~(\ref{s_virtuals}b) to $\bn^2=0$.
The diagram~(\ref{s_virtuals}c) and its Hermitian conjugate give
\begin{align}\label{s_virtuals_c}
\phi_1^{(\ref{s_virtuals}c)}&=
-2ig^2 C_F \d^{(2)}(\vec k_{n\perp}) \mu^{2 \eps}
\int \frac{d^d k}{(2 \pi)^d} \frac{1}{[k^+-i\d^+] [k^-+i\d^-] [k^2+i0]} +h.c.
\nn \\
&=
- \frac{\alpha_s C_F}{2\pi}
\d^{(2)}(\vec k_{n\perp})
\left[\frac{2}{\veuv^2}-\frac{2}{\veuv}\ln\frac{\d^+\d^-}{\mu^2}+
\ln^2\frac{\d^+\d^-}{\mu^2}+\frac{\pi^2}{2}\right]\, .
\end{align}

The virtual part of the TMDPDF at ${\cal O}(\as)$ using the relation in Eq.~(\ref{regul_DeltaDY}) is
\begin{align}
\label{eq:jvn}
 j_{n1}^v &= - \frac{1}{2} \hat f_{n1}^{(\ref{n_virtuals}a)} + \hat f_{n1}^{(\ref{n_virtuals}c)}
 - \frac{1}{2} \d(1-x) \phi_1^{(\ref{s_virtuals}c)}
=
\frac{\alpha_s C_F}{2 \pi}
\d(1-x)\d^{(2)}(\vec k_{n\perp})
\left[
\frac{1}{\veuv^2} + \frac{1}{\veuv} \left( \frac{3}{2} + \ln\frac{\mu^2}{Q^2} \right)
\right.
\nn\\
&\left.
-\frac{3}{2}\ln\frac{\D}{\mu^2} - \frac{1}{2}\ln^2\frac{\D^2}{Q^2\m^2} + \ln^2\frac{\D}{\m^2}
+ \frac{7}{4} - \frac{\pi^2}{3}\right]\,,
\end{align}
where we have set $\D^+=\D^-=\D$. Analogously the anticollinear one is
\begin{align}
\label{eq:jvnb}
j_{\bn 1}^v &=
\frac{\alpha_s C_F}{2 \pi}
\d(1-z)\d^{(2)}(\vec k_{\bn\perp})
\left[
\frac{1}{\veuv^2} + \frac{1}{\veuv} \left( \frac{3}{2} + \ln\frac{\mu^2}{Q^2} \right)
\right.
\nn\\
&\left.
-\frac{3}{2}\ln\frac{\D}{\mu^2} - \frac{1}{2}\ln^2\frac{\D^2}{Q^2\m^2} + \ln^2\frac{\D}{\m^2}
+ \frac{7}{4} - \frac{\pi^2}{3}\right]\,.
\end{align}
As mentioned earlier, individual contributions to $j_{n1(\bn 1)}^v$ have mixed UV and IR divergences, however $j^v_{n1(\bn 1)}$ itself is free from them.

%%%%%%%%%%%%%%%%%%%%%%%%%%%%%%%%%
\subsection{Real Diagrams}
\label{sec:real}
%%%%%%%%%%%%%%%%%%%%%%%%%%%%%%%%%

%%%%%%%%%%%%%%%%%%%%%%%%%%%%%%%%%%%%FIGURE
\begin{figure}
\begin{center}
\includegraphics[width=\textwidth]{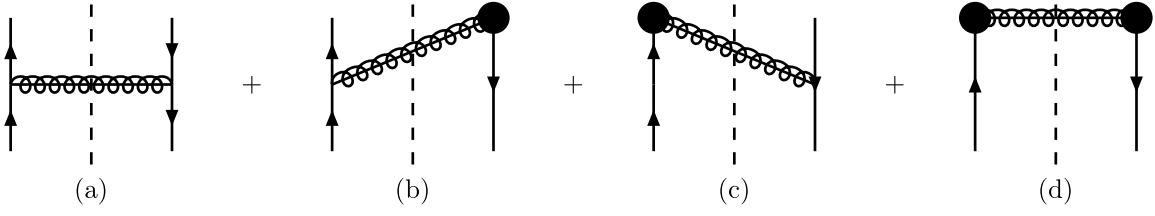}
\end{center}
\caption{\it Real gluon contributions for $\hat f_{n(\bn)}$.
\label{n_reals}}
\end{figure}
%%%%%%%%%%%%%%%%%%%%%%%%%%%%%%%%%%%%ENDFIGURE

\begin{figure}
\begin{center}
\includegraphics[width=\textwidth]{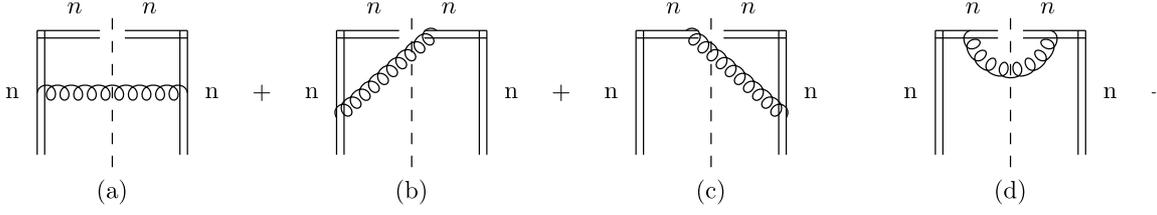}
\end{center}
\caption{\it Real gluon contributions for the Soft function.
\label{s_reals}}
\end{figure}

The relevant diagrams  for the real part of $j_n$ are shown in  figs.~(\ref{n_reals}) and~(\ref{s_reals}). The diagram~(\ref{n_reals}a) gives
\begin{align}\label{eq:jn3a}
\hat f_{n1}^{(\ref{n_reals}a)}&=
2\pi g^2 C_F p^+ \int\frac{d^dk}{(2\pi)^d}
\d(k^2)\theta(k^+)\frac{2(1-\ve)|\vec k_\perp|^2}{[(p-k)^2+i\D^-] [(p-k)^2-i\D^-]}
\nn\\
&\times
\d\le((1-x)p^+-k^+\ri) \d^{(2)}(\vec k_\perp+\vec k_{n\perp})
\nn\\
&=
\frac{2\a_s C_F}{(2\pi)^{2-2\ve}}
(1-\ve)(1-x)
\frac{|\vec k_{n\perp}|^2}{\left||\vec k_{n\perp}|^2-i\D^-(1-x)\right|^2}\,,
\end{align}
where the $\ve$-dependence and the deltas are kept for later convenience. The $\ve$ will be needed when transforming to impact parameter space in pure DR, and hence putting all deltas to zero. The deltas will be needed when going to impact parameter space in 4 dimensions, using them to regulate $k_\perp\to 0$. The sum of diagram~(\ref{n_reals}b) and its Hermitian conjugate~(\ref{n_reals}c) is
\begin{align}\label{eq:jn3b3c}
\hat f_{n1}^{(\ref{n_reals}b+\ref{n_reals}c)}&=
-4\pi g^2 C_F p^+  \int\frac{d^dk}{(2\pi)^d}
\d(k^2)\theta(k^+)\frac{p^+-k^+}{[k^++i\d^+][(p-k)^2+i\D^-]}
\nn\\
&\times
\d\le((1-x)p^+-k^+\ri) \d^{(2)}(\vec k_\perp+\vec k_{n\perp}) + h.c.
\nn\\
&=
\frac{2\alpha_s C_F}{(2\pi)^{2-2\ve}}
\left[\frac{x}{(1-x)+i\d^+/p^+}\right]
\left[
\frac{1}{|\vec k_{n\perp}|^2-i\D^-(1-x)}
\right]
+ h.c.
%\nn\\
%&=
%\frac{2\alpha_s C_F}{(2\pi)^{2-2\ve}} \frac{1}{|\vec k_{n\perp}|^2}
%\left[\frac{2x}{(1-x)_+} - 2\delta(1-x) \ln\frac{\d^+}{p^+}\right]
\,,
\end{align}
Diagram~(\ref{n_reals}d) is zero, since it is proportional to $\bn^2=0$.

%%%%%%%%%%%%%%%%%%%%%%%%%%%%%%%%%%%%FIGURE
%%%%%%%%%%%%%%%%%%%%%%%%%%%%%%%%%%%%ENDFIGURE
For the real emission of soft gluons, the diagrams~(\ref{s_reals}a) and~(\ref{s_reals}d) are zero, since they are proportional to $n^2=0$ and $\bn^2=0$ respectively.
The diagram~(\ref{s_reals}b) and its Hermitian conjugate~(\ref{s_reals}c) give
\begin{align}\label{eq:jn4b4c}
\phi_1^{(\ref{s_reals}b+\ref{s_reals}c)}&=
-4\pi g^2 C_F  \int\frac{d^dk}{(2\pi)^d}
\d^{(2)}(\vec k_\perp+\vec k_{n\perp}) \d(k^2)\theta(k^+) \frac{1}{[k^++i\d^+][-k^-+i\d^-]} + h.c.
\nn\\
&=
-\frac{4\a_s C_F}{(2\pi)^{2-2\ve}}
\frac{1}{|\vec k_{n\perp}|^2-\d^+\d^-} \ln\frac{\d^+\d^-}{|\vec k_{n\perp}|^2}\,.
%=
%-\frac{4\a_s C_F}{(2\pi)^{2-2\ve}}
%\frac{1}{|\vec k_{n\perp}|^2} \ln\frac{\d^+\d^-}{|\vec k_{n\perp}|^2}\,.
\end{align}
To get the real part of the TMDPDF we need to put together all the contributions above into Eq.~(\ref{jn}). In order to do that we set the deltas to zero unless they are necessary to regulate any divergence, i.e., we keep them in the logs but not for the limit $|k_\perp|\to 0$. The result, setting $\D^\pm=\D$, is
\begin{align}
\label{eq:jr}
j_{n1}^r&= \hat f_{n1}^{(\ref{n_reals}a)} + \hat f_{n1}^{(\ref{n_reals}b+\ref{n_reals}c)}
- \frac{1}{2} \delta(1-x) \phi_1^{(\ref{s_reals}b+\ref{s_reals}c)}
\nn\\
&=
 \frac{2\as C_F}{(2\pi)^{2-2\ve}} \frac{1}{|\vec k_{n\perp}|^2} \left[
(1-\ve)(1-x) + \frac{2 x}{(1-x)_+}
+ \delta(1-x) \ln\frac{Q^2}{|\vec k_{n\perp}|^2}
\right]\,,
\end{align}
where the $\ln\D$ has cancelled in the combination of Eqs.~(\ref{eq:jn3b3c}) and~(\ref{eq:jn4b4c}), using the relation~(\ref{eq:distributions}) to express the first one in terms of distributions.
A similar result is obtained for the anticollinear TMDPDF
\begin{align}
\label{eq:jrnb}
j_{\bn 1}^r &=
\frac{2\a_s C_F}{(2\pi)^{2-2\ve}} \frac{1}{|\vec k_{\bn\perp}|^2}\left[
(1-\ve)(1-z) + \frac{2 z}{(1-z)_+}
+ \delta(1-z) \ln\frac{Q^2}{|\vec k_{\bn\perp}|^2}
\right]\, ,
\end{align}
where the superscript $r$ stands for ``real'' gluon contributions. Our results for the real gluon emission agree with the ones in \cite{Ji:2004wu,Becher:2011xn}.

%%%%%%%%%%%%%%%%%%%%%%%%%%%%%%%%%%%%%%%%
\subsection{Where Are the Rapidity Divergencies?}
\label{sec:rapidity}
%%%%%%%%%%%%%%%%%%%%%%%%%%%%%%%%%%%%%%%%

As can be seen from Eqs.~(\ref{eq:jr}-\ref{eq:jrnb}) the contribution to the TMDPDF from real gluon emission does not include any logarithmic dependence on $\delta$'s after a fine cancelation of those logarithms between the naive collinear and the square root of the soft contribution.
 Clearly however this regulator is needed to calculate individual Feynman diagrams in order to avoid light-cone singularities. Now let us consider  the contribution from virtual diagrams given 
in Eqs. (\ref{eq:jvn}-\ref{eq:jvnb}). 
In obtaining those results there is also a fine cancelation of the pieces with mixed divergences appearing in both terms of the soft and naive collinear contributions.
However the remaining $\Delta$-dependence can be safely interpreted as pure IR one, namely soft or collinear, and not as rapidity divergence. 
The reason is simple. 
In Appendix~\ref{sec:app2} we have calculated the quark form factor in full QCD using the same $\delta$-regulator. 
Since QCD has no rapidity divergences (or in other words there are no Wilson lines) then all the $\D$-dependence is pure IR. Those IR terms (see Eq.~(\ref{eq:currentdy})) are exactly the ones appearing in Eqs.~(\ref{eq:jvn}-\ref{eq:jvnb}). Among other things this fact allows for QCD to be matched onto SCET at the higher scale $Q$ thus obtaining a matching coefficient, $H$, which is free from any IR dependence. Moreover the logarithms of $\Delta$'s in the virtual contribution need not to be resummed. So we conclude  that our TMDPDF is free from any rapidity divergences.
  
To strengthen the above statement we mention that one can also obtain a $\D$-free virtual contributions by simply calculating them in pure DR, as we show below.
 Also in that case there is a fine cancelation of mixed poles (UV and IR) between the naive collinear and the square root of the soft function 
and the virtual contributions contain both UV poles and IR ones. 
The IR ones will again exactly match the ones of QCD for the quark form factor.
Then the complete first order result for the  TMDPDF can be obtained without any $\d$-dependence. In other words it is free from any rapidity divergences.
It is interesting to notice that also when using the regulator introduced in \cite{Chiu:2012ir} (the $\nu$-regulator) our observation above is still valid.
 In \cite{Chiu:2012ir} the zero-bin contribution is zero to all orders in perturbation theory. However the soft function defined in that work, which at the operator level is identical to ours, is non-vanishing and $\nu$-dependent. In this case we resort to our defining formula for the TMDPDF, namely  Eq.~(\ref{eq:jtmdgeneral}). 
Taking the results from \cite{Chiu:2012ir} (Eqs.~(5.50, 5.51) for collinear and Eq.~(5.62) for the soft contribution) one can easily verify that the TMDPDF, calculated with $\nu$-regulator is also independent of the rapidity divergence regulator $\nu$. Although the above discussion relies on first order calculations in $\as$, it holds to all orders in perturbation theory as we explain below.

The full QCD calculation of the partonic cross section can be performed while staying on-the-light-cone and using pure DR to regularize both the UV and IR divergences. When matching QCD and SCET at the higher scale $Q$ the hard matching coefficient should be independent of any IR regulator to all orders in perturbation theory (as long as the assumption that SCET captures all IR behavior of QCD is still intact).\footnote{
In \cite{Ji:2004wu,Chen:2006vd} the QCD matrix element was calculated on-the-light-cone, while the SCET ones where calculated off-the-light-cone. Thus, the hard matching coefficient is dependent on the off-the-light-coneness parameter $\rho$.
}
Given the relation in Eq.~(\ref{step42}), it is clear that $j_{n(\bn)}$ should also be free from any rapidity divergence regulator since the full QCD result is.

Consider the virtual contribution to the TMDPDF given in Eq.~(\ref{eq:jvn}) calculated with the $\d$-regulator. Since the real gluon contribution (see Eq.~(\ref{eq:jr})) is free from that regulator, then the TMDPDF does depend on it and this dependence persists also at higher orders in perturbation theory. However the important thing to notice is that the anomalous dimension $\g_n$ of the TMDPDF and its matching coefficient $\tilde C_n$ onto the PDF, which are the only perturbatively calculable physical quantities relevant to the \emph{hadronic} TMDPDF (in the kinematical regime we are interested in), are still $\d$-free. This is due to the fact that in Eq.~(\ref{eq:jvn}) we managed to separate the UV from  all the IR divergences including the light-cone ones. In Section~\ref{sec:tmdpdftopdf} and in Appendix~\ref{sec:app} we show this explicitly to first order in $\as$. The derivation of the anomalous dimension of the TMDPDF is given in Secs.~\ref{sec:AD} and~\ref{sec:adpuredr}.

As mentioned in the Introduction, one of the prominent methods to regularize light-cone singularities is to go off-the-light-cone where we tilt the light-cone vectors: $(n,\bn) \to (v, \bv)$, where $v^2 \neq 0$, $\bv^2 \neq 0$. The diagrams with real gluon emission that have light-cone singularities and need to be regularized are~(\ref{n_reals}b,\ref{n_reals}c,\ref{s_reals}b,\ref{s_reals}c). From the naive collinear contribution we get
\begin{align}\label{eq:n_reals_ji}
\hat f_{n1}^{(\ref{n_reals}b+\ref{n_reals}c)}&=
\frac{2\as C_F}{(2\pi)^{2-2\ve}} \frac{1}{|\vec k_{n\perp}|^2} \left[
\frac{2x}{(1-x)_+} + \d(1-x)\ln\frac{Q^2 \rho}{|\vec k_{n\perp}|^2} \right]\,,
\end{align}
and from the soft contribution we have
\begin{align}\label{eq:s_reals_ji}
\phi_1^{(\ref{s_reals}b+\ref{s_reals}c)}&=
\frac{2\as C_F}{(2\pi)^{2-2\ve}} \frac{1}{|\vec k_{n\perp}|^2} \ln\rho^2\,,
\end{align}
where $\rho = (\bv^-/\bv^+) = (v^+/v^-)$ and goes to infinity in the light-cone limit.
Eqs.~(\ref{eq:n_reals_ji}-\ref{eq:s_reals_ji}) are consistent with the results found in \cite{Ji:2004wu}.
Combining these two results according to Eq.~(\ref{eq:jr}) with $\hat f_{n1}^{(\ref{n_reals}a)}$ given in Eq.~(\ref{eq:jn3a}), we get again the result in the second line of Eq.~(\ref{eq:jr}), where all the $\rho$-dependence cancels in exactly the same manner as the $\d$-dependence was canceled. The last observation differs from the one found by Collins \cite{Collins:2011zzd}. The reason is that in Collins' work, the contribution from the collinear matrix element to the TMDPDF is treated while staying on-the-light-cone and only the soft function is ``tilted'' (see Eq.~(10.136) in \cite{Collins:2011zzd}). In \cite{Ji:2004wu} (see also \cite{Chen:2006vd}) all matrix elements, soft and collinear, are taken off-the-light-cone in a consistent manner. In this case, as we have shown, the $\rho$-dependence cancels in the TMDPDF.

The authors in~\cite{Chiu:2012ir} have a different definition of the TMDPDF than ours in the sense that the soft function is not included in that definition. 
Moreover, since they identify the collinear matrix element as the TMDPDF, it contains rapidity divergencies, which are interpreted as UV ones (because of the regulator they use), 
and are then resummed using usual RG techniques.

%%%%%%%%%%%%%%%%%%%%%%%%%%%%%%%%%
\subsection{Anomalous Dimension of the TMDPDF $j_{n(\bn)}$}
\label{sec:AD}
%%%%%%%%%%%%%%%%%%%%%%%%%%%%%%%%%

In order to compute the anomalous dimension of the TMDPDF to ${\cal O}(\alpha_s)$ one needs to consider only its virtual contributions since, as we have seen in the previous sub-section, all real contributions are UV-finite. From Eq.~(\ref{eq:jvn}) the counterterm for the $n$-collinear TMDPDF is
\begin{align}
{\cal Z}_n &= 1- \frac{\alpha_s C_F}{2\pi}
\left[ \frac{1}{\veuv^2} + \frac{1}{\veuv} \left( \frac{3}{2} + \ln\frac{\mu^2}{Q^2} \right)
\right],
\label{eq:zn}
\end{align}
and the  corresponding anomalous dimension is
\begin{align} \label{g_n}
\g_n &= \frac{d \ln {\cal Z}_n}{d\ln \m} =
%\frac{1}{Z} \frac{\pd Z}{\pd \ln\m} + \frac{1}{Z}\frac{\pd Z}{\pd \a_s}\frac{\pd \a_s}{\pd \ln\m}=
\frac{1}{{\cal Z}_n} \frac{\pd {\cal Z}_n}{\pd \ln\m} + \frac{1}{{\cal Z}_n} \frac{\pd {\cal Z}_n}{\pd \a_s}(-2\ve\a_s+O(\a_s^2))
\nn\\
\g_{n1} &=  \frac{\alpha_s C_F}{2\pi} \left[ 3 + 2\ln\frac{\mu^2}{Q^2} \right]\,.
\end{align}
For the $\bn$-collinear sector we have, analogously, ${\cal Z}_{\bar n}$, from which we get
\begin{align}\label{g_bn}
\g_{\bn 1} &= \frac{d \ln {\cal Z}_{\bar n}}{d\ln \m} %=
%\frac{1}{Z} \frac{\pd Z}{\pd \ln\m} + \frac{1}{Z}\frac{\pd Z}{\pd \a_s}\frac{\pd \a_s}{\pd \ln\m}=
%\frac{1}{{\cal Z}_{\bar n}} \frac{\pd \cal{Z}_{\bar n}}{\pd \ln\m} + \frac{1}{\cal{Z}_{\bar n}}
 %\frac{\pd {\cal Z}_{\bar n}}{\pd \a_s}(-2\ve\a_s+O(\a_s^2))
%\nn\\ &
 =  \frac{\alpha_s C_F}{2\pi} \left[ 3 + 2\ln\frac{\mu^2}{Q^2} \right]\,.
\end{align}

Now we compare our result for  the anomalous dimension with the one of Collins~\cite{Collins:2011zzd}
\footnote{The factor of $2$ difference between Eqs.~(\ref{g_n}, \ref{g_bn}) and Eq.~(\ref{g_collins}) in the logarithmic term is due to different conventions for the light-cone components.}
\begin{align}\label{g_collins}
\g_{n1,Collins} &=
\frac{\alpha_s C_F}{2\pi} \left[ 3 + 2\ln\frac{\mu^2}{2 (p^+)^2 e^{-2y_n}} \right],
\end{align}
where $y_n$ is a measure of ``off-light-coneness'' which sets a lower bound on the rapidity of gluons. Collins uses tilted vectors along the trajectories of the incoming hadrons $n_t$ and $\bn_t$ which are parameterized as $n_t=(1,-e^{-2y_n},0_\perp)$ and $\bn_t=(-e^{2y_n},1,0_\perp)$. Thus it is clear that in order, for example, to recover the light-cone limit one has to consider \emph{two} different limits: $y_n \rightarrow \infty$ for $n_t$ and $y_n \rightarrow -\infty $ for $\bn_t$. This means that the light-cone limit in $\g_{n1,Collins}$ is unattainable.
Moreover, in Ref.~\cite{Collins:2011zzd}, the $\zeta$-dependence cannot be eliminated by any single natural choice. In our case, however, we see that by setting one regulator, $\D=\D^+=\D^-$, the anomalous dimension becomes independent of that regulator and it is free from rapidity divergences.

Moreover, the evolution of the TMDPDF can be governed only by the $\m^2$-evolution and no additional parameters are needed.

In Sec.~\ref{sec:resummation} we give the AD of the TMDPDF at second order in $\as$.

%%%%%%%%%%%%%%%%%%%%%%%%%%%%%%%%%%%%%%%%
\subsection{Anomalous Dimension On-The-Light-Cone in Dimensional Regularization}
\label{sec:adpuredr}
%%%%%%%%%%%%%%%%%%%%%%%%%%%%%%%%%%%%%%%%

It is possible to calculate the anomalous dimension of the TMDPDF using pure DR as in Ref.~\cite{Idilbi:2007ff}.
At one-loop, we need only to consider the virtual contributions given in diagrams~(\ref{n_virtuals}a),~(\ref{n_virtuals}c)  and~(\ref{s_virtuals}c). All the rest vanish identically due to light-like Wilson lines.
For diagram~(\ref{n_virtuals}a) (without its Hermitian conjugate) we have
\begin{equation}
\hat f_{n1}^{(\ref{n_virtuals}a)}=
\frac{\alpha_s C_F}{4\pi}
\d(1-x)\d^{(2)}(\vec k_{n\perp})
\left(\frac{1}{\veuv}-\frac{1}{\veir}\right)\,\, ,
\end{equation}
and for diagram~(\ref{n_virtuals}c),
\begin{equation}
\label{me1}
\hat f_{n1}^{(\ref{n_virtuals}c)}=
\frac{\alpha_s C_F}{4\pi}
\d(1-x)\d^{(2)}(\vec k_{n\perp})
\left(\frac{\mu^2}{-\kappa (p^+)^2}\right)^\varepsilon\left[ \frac{1}{\veir}\left(\frac{2}{\veuv}-\frac{2}{\veir}\right)+\left(\frac{2}{\veuv}-\frac{2}{\veir}\right)\right]\,\,.
\end{equation}
Notice that in this regularization scheme the energy scale inside the logs is fixed noting that $p^+$ is the
only relevant scale in the virtual part of the TMDPDF. Thus the scale inside the logs is equal to $-\kappa (p^+)^2$ where $\kappa = Q^2/(p^+)^2$ and it is required to remove the dimensional ambiguity in integrals of the form: $\int_0^\infty dt~t^{-1-\varepsilon}$.
The soft function, diagram~(\ref{s_virtuals}c), gives
\begin{equation}
\label{me2}
\phi_1^{(\ref{s_virtuals}c)}=
- \frac{\alpha_s C_F}{2\pi}
\d^{(2)}(\vec k_{n\perp})
\left(\frac{\mu^2}{-\kappa (p^+)^2}\right)^\varepsilon 2\left[ \frac{1}{\veuv}-\frac{1}{\veir}\right]^2\,\,.
\end{equation}
Taking into account the Hermitian conjugate diagrams, the total virtual contribution to the TMDPDF is
\begin{equation}
\label{eq:jvdimreg}
j_{n1}^v=
\frac{\alpha_s C_F}{2\pi}
\d(1-x)\d^{(2)}(\vec k_{n\perp})
\left[ \frac{1}{\veuv^2}-
\frac{2\ln \left(\frac{\kappa (p^+)^2}{\mu^2}\right)-3}{2\veuv}-
\frac{1}{\veir^2}+\frac{2\ln \left(\frac{\kappa (p^+)^2}{\mu^2}\right)-3}{2\veir}\right]\,\,.
\end{equation}
From the result for $j_{n1}^v$ one can easily identify the counter-term ${\cal Z}_n$ needed to cancel the UV divergences. Defining $\g_n=\frac{d\ln {\cal Z}_n}{d\ln \mu}$ one gets
\begin{equation}
\g_{n1}=\frac{\alpha_s C_F}{2\pi}\left[3+2\ln\frac{\mu^2}{\kappa (p^+)^2}\right]
=
\frac{\alpha_s C_F}{2\pi}\left[3+2\ln\frac{\mu^2}{Q^2}\right]\,,
\end{equation}
which agrees with Eq.~(\ref{g_n}).

In Eq.~(\ref{eq:jvdimreg}) we again notice that there are no mixed UV and IR divergences as was observed in Eqs.~(\ref{eq:jvn}, \ref{eq:jvnb}). It should be noted that if one had subtracted the complete soft function from the collinear part (and not the square root of it) then there would be mixed UV and IR poles and those mixed poles would not cancel even after including the contribution from real gluon emission. This would definitely prevent such quantity from being an acceptable definition of TMDPDF.

%%%%%%%%%%%%%%%%%%%%%%%%%%%%%%%%
%%%%%%%%%%%%%%%%%%%%%%%%%%%%%%%%
\section{From TMDPDF to Integrated PDF}
\label{sec:tmdpdftopdf}
%%%%%%%%%%%%%%%%%%%%%%%%%%%%%%%%

We recall from Section~\ref{sec:fac} the OPE of the TMDPDF onto the PDF,
\begin{align}\label{eq:ope}
\tilde j_n(x;\vec b_\perp,Q,\m) = \int_x^1 \frac{dx'}{x'} \tilde C_n\le(\frac{x}{x'};b,Q,\m\ri)\, {\cal Q}_n(x';\m)\,,
\end{align}
where
\begin{align}
{\cal Q}_n(x;\m) = \frac{1}{2} \int \frac{dy^-}{2\pi} e^{-i\frac{1}{2}y^-xp^+}
\sandwich{p}{\bar \chi_n(0^+,y^-,\vec 0_\perp) \frac{\bnslash}{2}\chi_n^\dagger(0^+,0^-,\vec 0_\perp)}{p}
|_\textrm{zb~ included}\,.
\end{align}
In this section we compute $\tilde C_n$ to first order in $\as$ and also establish to the same order the following relation in $d=2-2\ve$
\begin{align}\label{tmdpdftopdfall}
\m^{2\ve}\int d^d\vec k_{n\perp}\, j_{n(\bn)}(x;\vec k_{n\perp},Q,\m) &= {\cal Q}_n(x;\m)\,,
\end{align}
which is expected to hold to all orders in perturbation theory for bare quantities. We comment more on this below.

In impact parameter space the virtual and real parts of the TMDPDF are obtained from Eqs.~(\ref{eq:jvdimreg}, \ref{eq:jr}) by Fourier transformation. The results are
\begin{align}
\tilde j_{n1}^v(x;\vec b_\perp,Q,\m) &=
\frac{\alpha_s C_F}{2 \pi}
\d(1-x)
\left[
\frac{1}{\veuv^2} + \frac{1}{\veuv} \left( \frac{3}{2} + \ln\frac{\mu^2}{Q^2} \right)
- \frac{1}{\veir^2} - \frac{1}{\veir} \left( \frac{3}{2} + \ln\frac{\mu^2}{Q^2} \right)
\right]\,,
\nn\\
\tilde j_{n1}^r(x;\vec b_\perp,Q,\m)&=
\frac{\a_s C_F}{2\pi}\left[
\left( {\cal P}_{q/q}(x)-\d(1-x)\left(\frac{3}{2} + \ln\frac{\m^2}{Q^2}\right) \right)
\left(-\frac{1}{\veir}-L_T \right) \right.
\nn\\
&\left.
+(1-x) +\d(1-x) \Big(\frac{1}{\veir^2}-\frac{1}{2}L_T^2 -\frac{\pi^2}{12}\Big)
\right]\,,
\end{align}
where $L_T = \ln(\m^2b^2e^{2\g_E}/4)$ and
\begin{align}
{\cal P}_{q/q} = \left( \frac{1+x^2}{1-x} \right)_+ =
\frac{1+x^2}{(1-x)_+} + \frac{3}{2}\d(1-x) =
\frac{2x}{(1-x)_+} + (1-x) + \frac{3}{2}\d(1-x)\,,
\end{align}
which is the one-loop quark splitting function of a quark in a quark. We have used pure DR and dropped the $\D^-$ in the real diagrams. In Appendix~\ref{sec:app} this calculation is done while keeping the $\D^-$ to regulate $k_{n\perp}\to 0$ when going to the impact parameter space.

The complete renormalized TMDPDF in impact parameter space, to first order in $\as$, becomes
\begin{align}\label{eq:jips}
\tilde j_n(x;\vec b_\perp,Q,\m) &=
\d(1-x) + \frac{\alpha_s C_F}{2 \pi}\left\{
{\cal P}_{q/q}(x)\left(  -\frac{1}{\veir} - L_T \right) + (1-x)
\right.
\nn\\
&\left.
-\d(1-x)\left[
\frac{1}{2}L_T^2 - \frac{3}{2}L_T
+\ln\frac{Q^2}{\m^2}L_T + \frac{\pi^2}{12}
\right]
\right\}\,.
\end{align}
Given that to first order in $\as$ the renormalized PDF is
\begin{align}\label{eq:pdfir}
{\cal Q}_n(x;\m) &= \d(1-x)+ \frac{\alpha_s C_F}{2 \pi} {\cal P}_{q/q}(x) \left(-\frac{1}{\veir} \right)\,,
\end{align}
we extract from Eq.~(\ref{eq:ope}) the matching coefficient at first order in $\as$,
\begin{align}\label{coeff}
\tilde C_n(x;b,Q,\m) &=  \d(1-x) + \frac{\alpha_s C_F}{2 \pi}
\left[-{\cal P}_{q/q} L_T + (1-x)
\right.
\nn\\
&\left.
- \d(1-x)\left( \frac{1}{2}L_T^2 - \frac{3}{2} L_T + \ln\frac{Q^2}{\m^2}L_T
+\frac{\pi^2}{12}  \right) \right]\,.
\end{align}
At this stage it is worth noticing the appearance of $\ln(Q^2/\m^2)$ at the matching coefficient. From the above result, we can see that by a proper choice of the scale $\m=\m_I=(2e^{-\g_E}/b)$, we eliminate this logarithm since $L_T(\m_I)=0$. However at this order in perturbation theory this cancelation is accidental and it does not persist at higher orders. In subsection~\ref{sec:q2} we discuss the appearance of $\ln(Q^2/\m^2)$ at an arbitrary order in perturbation theory and how to handle them.

It has been a matter of debate whether the PDF can be obtained from the naive collinear matrix element, ${\hat f}_{n(\bn)}$,  by simple integration over the transverse momentum $\vec k_\perp$ conjugate to $b_\perp$ (see, e.g., \cite{Ji:2004wu,Collins:2011zzd,Cherednikov:2007tw} and  more recently \cite{Aybat:2011zv,Anselmino:2012aa}). In principle, and  as a consistency check of the partonic definitions of such quantities, this should be the case; however, and interestingly enough, this was never established before.
  Below we establish that even with the inclusion of the soft function in the definition of the TMDPDF we are able to recover the PDF in a straightforward manner. It should be mentioned that going from TMDPDF to PDF can only be obtained when considering bare quantities. In other words if we consider the renormalized TMDPDF and then integrate over the transverse momentum we cannot recover the renormalized PDF since the integration over the transverse momentum introduces new UV divergences that need to be renormalized in their own turn. Stated differently, it is the lack of interchangeability between integration and subtraction (of UV divergences) which prohibit the passage from renormalized TMDPDF to PDF. This is true whether the soft function is included or not, in the definition of the TMDPDF.
  
 To do so we need the following integrals in $d=2-2\ve$ and in $\overline{\textrm{MS}}$ scheme
\begin{align}
\m^{2\ve}\int d^d\vec k_{\perp} \frac{1}{|\vec k_\perp|^2} &=
\pi \left( \frac{1}{\veuv} - \frac{1}{\veir} \right)\,,
\nn\\
\m^{2\ve}\int d^d\vec k_\perp \frac{1}{|\vec k_\perp|^2} \ln|\vec k_\perp|^2 &=
\pi \left( \frac{1}{\veuv^2} - \frac{1}{\veir^2} \right) +
\pi \left( \frac{1}{\veuv} - \frac{1}{\veir} \right)\ln\frac{\m^2}{(2\pi)^2}\,,
\end{align}
from which we get
\begin{align}
\m^{2\ve}\int d^d\vec k_{n\perp}\, j_{n1}^v &=
\frac{\alpha_s C_F}{2 \pi} \d(1-x) \left[
\frac{1}{\veuv^2} + \frac{1}{\veuv} \left( \frac{3}{2} + \ln\frac{\mu^2}{Q^2} \right)
- \frac{1}{\veir^2} - \frac{1}{\veir} \left( \frac{3}{2} + \ln\frac{\mu^2}{Q^2} \right)
\right]\,,
\nn\\
\m^{2\ve}\int d^d\vec k_{n\perp}\, j_{n1}^r &=
\frac{\a_s C_F}{2\pi}
\left\{ \left[ \d(1-x)\ln\frac{Q^2}{\m^2} + {\cal P}_{q/q} - \frac{3}{2}\d(1-x)\right] \left( \frac{1}{\veuv} - \frac{1}{\veir} \right)
\right.
\nn\\
&\left.
- \d(1-x) \left( \frac{1}{\veuv^2} - \frac{1}{\veir^2} \right)
\right\}\,.
\end{align}
Taking the sum of the previous  equations one gets
\begin{align}\label{tmdpdftopdf1}
\m^{2\ve}\int d^d\vec k_{n\perp}\, j_n &= \d(1-x) + \frac{\a_s C_F}{2\pi} {\cal P}_{q/q} \left(\frac{1}{\veuv}- \frac{1}{\veir} \right)\,,
\end{align}
which is the PDF shown in Eq.~(\ref{eq:pdfir}). To the best of our knowledge this consistency check between the TMDPDF and the PDF is established for the first time. We should also mention that even without the inclusion of the soft function in the definition of the TMDPDF then one can repeat the above manipulations by considering only the naive collinear matrix element  ${\hat f}_{n(\bn)}$ and then recover the bare PDF by integrating over transverse momentum coordinates.

Now, from Eqs.~(\ref{eq:ope}) and~(\ref{tmdpdftopdfall}) we extract the following condition for the matching coefficient at the intermediate scale in momentum space
\begin{align}\label{eq:coeffeq}
\int d^d\vec k_{n\perp}\, C_n(x;\vec k_{n\perp},Q,\m) \equiv \int d^d\vec k_{n\perp}\, \sum_{i=0}^\infty \as^i C_{ni}(x;\vec k_{n\perp},Q,\m) &= \d(1-x)\,,
\nn\\
\int d^d\vec k_{n\perp}\, C_{ni}(x;\vec k_{n\perp},Q,\m) &= 0 \,,\quad \forall i>0 \,,
\end{align}
where $C_{n0}(x;\vec k_{n\perp},\m)=\d(1-x)\d^{(2)}(\vec k_{n\perp})$.
It should be stressed that the last equations apply to the UV-finite contributions of the integrals of $C_{ni}(x;\vec k_{n\perp},\m)$ over $k_{n\perp}$ since clearly those integrals are UV-divergent in momentum space.

The coefficient at ${\cal O}(\as)$ in Eq.~(\ref{coeff}), once we transform it back to momentum space, is
\begin{align}
C_{n1}(x;\vec k_{n\perp},Q,\m) &= \frac{C_F}{2\pi^2}\left\{ (2\pi)^{2\ve} \left[
\left( {\cal P}_{q/q} - \frac{3}{2}\d(1-x) - \ve(1-x) \right) \frac{1}{|\vec k_{n\perp}|^2}
+ \d(1-x)\frac{1}{|\vec k_{n\perp}|^2} \ln\frac{Q^2}{|\vec k_{n\perp}|^2}
\right] \right.
\nn\\
&\left.
+\left[ \left({\cal P}_{q/q} - \frac{3}{2}\d(1-x) \right)\frac{\pi}{\veir}
+ \d(1-x)\left(-\frac{\pi}{\veir^2} + \frac{\pi}{\veir}\ln\frac{Q^2}{\m^2} \right)\right] \d^{(2)}(k_{n\perp})
\right\}\,.
\end{align}
Integrating over $\vec k_{n\perp}$ one can easily see that all the IR poles cancel and the remaining contribution contains only single and double UV poles. Actually one gets
\begin{align}
\int d^d\vec k_{n\perp}\, C_{n1}(x;\vec k_{n\perp},Q,\m) = \frac{C_F}{2\pi} \left[
\left( {\cal P}_{q/q} - \frac{3}{2}\d(1-x) + \ln\frac{Q^2}{\m^2}\d(1-x) \right)\frac{1}{\veuv}
- \d(1-x)\frac{1}{\veuv^2}
\right]\,,
\end{align}
thus the UV-finite term is zero and Eq.~(\ref{eq:coeffeq}) is established for $i=1$.

%%%%%%%%%%%%%%%%%%%%%%%%%%%%%%%%
\section{$Q^2$-Dependence and Resummation}
\label{sec:q2}
%%%%%%%%%%%%%%%%%%%%%%%%%%%%%%%%

The matching coefficient $\tilde C_n$ is expected to live at the intermediate scale $q_T \sim 1/b$. However the appearance of $\ln(Q^2/\m^2)$ in $\tilde C_{n1}$, and higher powers of it in higher orders in perturbation theory, might indicate otherwise. Notice, for example, that the logarithms in Eq.~(\ref{coeff}) cannot be combined into a simple logarithm, unlike the case of threshold region in inclusive Drell-Yan or DIS \cite{Manohar:2003vb,Idilbi:2005ky}. In the threshold region the matching coefficient at the intermediate scale $\m_I$ is a function of only one logarithm, $\ln(\m_I^2/\m^2)$.
Nonetheless, from general arguments concerning the $\d$-regulator we can extract and exponentiate this $Q^2$-dependence in the TMDPDF itself, thus putting it under control to all orders in perturbation theory.

Working in pure DR and setting all scaleless integrals to zero, only real diagrams contribute to $\tilde j_n$. Then, we can express the logarithm of the TMDPDF in impact parameter space as
\begin{align}\label{eq:lntildej}
\ln\tilde j_n = \ln\tilde{\hat{f_n}} - \frac{1}{2}\ln\tilde\phi\,,
\end{align}
where
\begin{align}\label{eq:r}
\ln\tilde{\hat{f_n}} &= {\cal R}_n \left( x;\as,L_T,\ln\frac{\d^+}{p^+} = \ln\frac{\D}{Q^2}  \right)\,,
\nn\\
\ln\tilde\phi &= {\cal R}_\phi \left( \as,L_T,\ln\frac{\d^+\d^-}{\m^2}=\ln\frac{\D^2}{Q^2\m^2} \right)\,,
\end{align}
and we have set $\D^\pm=\D$. The need for $\d$-regulator to regulate rapidity divergencies in individual Feynman diagrams of $\tilde{\hat{f_n}}$ and $\tilde\phi$ introduces the logarithmic dependencies shown in Eq.~(\ref{eq:r}). Due to dimensional arguments and Lorentz invariance, those are the only possible combinations that can appear.

Since the PDF is zero in pure DR and the matching coefficient between the TMDPDF and the PDF does not depend on the IR regulator, we have
\begin{align}\label{eq:delta_indep}
\frac{d}{d\ln\D} \ln\tilde j_n = 0\,,
\end{align}
which implies that ${\cal R}_n$ and ${\cal R}_\phi$ must be linear in their last arguments. 
Thus we can write
\begin{align}\label{eq:lnj}
\ln\tilde j_n = \ln\tilde j_n^{sub} - D(\as,L_T) \left( \ln\frac{Q^2}{\m^2} + L_T \right)\,.
\end{align}
where we have introduced $L_T$ just to cancel the $\m^2$-dependence in the coefficient of $D$ which simplifies the RG equations of the TMDPDF.
The function $\ln\tilde j_n^{sub}$ is independent of $Q^2$ and all the $Q^2$-dependence appears explicitly only in the $\ln(Q^2/\m^2)$.
Hence, we can extract all the $Q^2$-dependence from the TMDPDF and exponentiate it,  putting it under control and building what we will call the ``$Q^2$-factor'' hereafter.

We believe that the linearity in $\ln(Q^2/\m^2)$ can be extracted without relying on a particular scheme of regularization, but based on general arguments concerning the rapidity divergencies.
As we have shown in Sec.~\ref{sec:TMDPDF} to first order in $\as$, the TMDPDF is free from rapidity divergencies, since all the $\D$-dependence that remains exactly matches the IR contribution of full QCD. Then, although our $\d$-regulator does not differentiate the origin of the divergencies that it regulates, i.e., it encodes both the IR (soft and collinear) and rapidity divergencies, actually one could use another regulator that makes this distinction manifest. For instance the $\nu$-regulator introduced in~\cite{Chiu:2012ir}.

Now, if we denote by $\n$ the parameter that regulates only the rapidity divergencies (and using a different ones for the IR), then we believe that, based on the ${\cal O}(\as)$ calculation, the functional dependence of $\ln\tilde{\hat{f_n}}$ and $\ln\tilde\phi$ on $\n$ should be to all orders
\begin{align}\label{eq:r}
\ln\tilde{\hat{f_n}} &\longrightarrow \ln\frac{\n^2}{Q^2}\,,
\nn\\
\ln\tilde\phi &\longrightarrow \ln\frac{\n^2}{\m^2} = \ln\frac{\n^2}{Q^2} + \ln\frac{Q^2}{\m^2} \,,
\end{align}
where we have taken $p^+=\bp^-=Q$.

Since we know that the TMDPDF is free from rapidity divergencies, thus one can write
\begin{align}
\frac{d}{d\ln\n} \ln\tilde j_n = 0\,,
\end{align}
regardless on how the IR divergencies were regulated. And this equation again implies that ${\cal R}_n$ and ${\cal R}_\phi$ must be linear in the logs of $\n$, which automatically leads to Eq.~(\ref{eq:lnj}) and the extraction of $Q^2$ to all orders in perturbation theory into the $Q^2$-factor.

Using Eq.~(\ref{eq:ope}) the TMDPDF can be written as
\begin{align}\label{eq:q2factor}
\tilde j_n(x;\vec b_\perp,Q,\m) &= \left( \frac{Q^2 b^2 e^{2\g_E}}{4} \right)^{-D(\as,L_T)} \tilde {\cal C}_n(x;\vec b_\perp,\m)
\otimes {\cal Q}_n(x;\m)\,,
\end{align}
where
\begin{align}
\tilde {\cal C}_n(x;\vec b_\perp,\m)& =
\delta(1-x)+\frac{\alpha_s C_F}{2 \pi}\left[-{\cal P}_{q/q} L_T + (1-x) - \delta(1-x)
\left(-\frac{1}{2}L_T^2-\frac{3}{2}L_T+\frac{\pi^2}{12}\right)\right]\,.
\end{align}
The important thing to notice is that all the $Q^2$-dependence in the TMDPDF is exponentiated to all orders in perturbation theory where the exponent $D$ is perturbatively calculable and $\tilde {\cal C}_n$ is $Q^2$-independent.
Notice also that Eq.~(\ref{eq:q2factor}) refers to one single TMDPDF, and not to the product of both as in~\cite{Becher:2010tm}.

Given the renormalization group invariance of the hadronic tensor $\tilde M$ in impact parameter space,
\begin{align}
\tilde M = H(Q^2/\m^2)\, \tilde j_n(x;\vec b_\perp,Q,\m)\, \tilde j_\bn(z;\vec b_\perp,Q,\m)\ ,
\end{align}
(see  also Eq.~(\ref{eq:hadten}) below), we can establish the following relation between the AD of the hard matching coefficient, $\g_H$, and the one of the TMDPDFs, $\g_{n(\bn)}$,
\begin{align}\label{eq:ads}
\g_H &= -\g_n - \g_\bn = -2\g_n\,,
\end{align}
where $\g_n = \g_\bn$ and
\begin{align}
\label{eq:ads1}
\g_H = \frac{d\ln H}{d\ln\m}\,,
\quad
\g_{n(\bn)} = \frac{d\ln\tilde j_{n(\bn)}}{d\ln\m}\,.
\end{align}
The AD of the hard matching coefficient is linear in $\ln(Q^2/\m^2)$ to all orders in perturbation theory \cite{Manohar:2003vb, Korchemsky:1987wg},
\begin{align}\label{eq:adhard}
\g_H &= A(\as)\,\ln\frac{Q^2}{\m^2} + B(\as)\,,
\end{align}
where $A(\as)$  and $B(\as)$ are perturbatively calculable and are known up to third order in $\as$. Thus we get
\begin{align}
\g_n &= -\frac{1}{2}A(\as)\,\ln\frac{Q^2}{\m^2} - \frac{1}{2} B(\as)\,.
\end{align}

Applying RG invariance to the cross section, and the fact that $A(\as)=2\Gamma_{\rm cusp}(\as)$ to all orders in perturbation theory, we get
\begin{align}
\frac{d D(\as,L_T)}{d\ln\m} = \G_{\rm cusp}(\as)\,.
\end{align}
The perturbative expansion of $D$ is
\begin{align}
D(\as,L_T) = \sum_{n=1}^\infty d_n(L_T) \left( \frac{\as}{4\pi} \right)^n\,,
\end{align}
where $d_1(L_T)$ can be straightforwardly extracted from Eq.~(\ref{eq:jips}) and it is: $d_1(L_T)=2C_F L_T$. $d_2(L_T)$ can be read off from the result in~\cite{Becher:2010tm} by taking half of their result for $d_2^q(L_T)$. The factor of half results from the fact that we are considering only one collinear sector rather than a combination of two. Thus
\begin{align}
d_2(L_T)=\frac{\G_0 \b_0}{4}L_T^2 + \frac{1}{2}\G_1 L_T + C_F C_A \left(\frac{404}{27}-14\z_3\right) - \left(\frac{112}{27}\right)C_F T_F n_f\,,
\end{align}
where we have used the following expansions of the cusp AD and the beta function $\b(\as) = d\as/d\ln\m$
\begin{align}
\G_{\rm cusp}(\as) = \sum_{n=1}^\infty \G_{n-1} \left( \frac{\as}{4\pi} \right)^n \,, \quad\quad
\b(\as) = -2\as \sum_{n=1}^\infty \b_{n-1} \left( \frac{\as}{4\pi} \right)^n\,.
\end{align}

It is worthwhile at this stage to compare our analysis for the $Q^2$ dependence of the TMDPDF with the one in \cite{Becher:2010tm}. There are two major differences: the first is related to the origin of this hard-scale dependence  and the second is to which quantities it contributes. In \cite{Becher:2010tm} the emergence of  the $Q^2$ at the intermediate scale is attributed to the so-called ``collinear anomaly'' resulting from the use of the analytic regularization scheme. This scheme breaks the symmetries of the classical SCET Lagrangian and thus it can mediate an interaction between two distinct and otherwise decoupled two collinear sectors. In this way the collinear anomaly compensates for the absence of a soft function in the factorization theorem for the $q_T$-dependent DY process obtained in \cite{Becher:2010tm}. Moreover this collinear anomaly appears only in a product of two transverse-momentum-dependent matrix elements. In this product all dependence on the analytic regularization parameters ($\alpha$ and $\beta$) is canceled however the $Q^2$ dependence emerges. As a result one can indeed recover the full QCD result for the relevant hadronic tensor. In our case the $Q^2$-factor appears for each one of the \emph{individual} TMDPDFs and it is due to the existence  of the soft function in the definition of the TMDPDF. We also mention that in the analytic regularization scheme the soft function vanishes at any arbitrary order in perturbation theory due to scaleless integrals so, trivially, only the tree level contribution survives. If one adopts this scheme in our definition of the TMDPDF then the remaining naive collinear contribution becomes ill-defined as argued in \cite{Becher:2010tm}. Turning this question around we might consider the factorization theorem in  \cite{Becher:2010tm} but with a regularization scheme in which the soft function introduced here does not vanish. In any such scheme (whether it is the $\d$-regulator, non-zero offshellness, massive quarks and gluons or even going off-the-light-cone) the factorization theorem in \cite{Becher:2010tm} will not reproduce the full QCD result.

%%%%%%%%%%%%%%%%%%%%%%%%%%%%%%%%
\subsection{Resummation}
\label{sec:resummation}
%%%%%%%%%%%%%%%%%%%%%%%%%%%%%%%%
\begin{figure}
\begin{center}
%\includegraphics[width=0.3\textwidth]{dy_small_qt}
%\quad\quad\quad
\includegraphics[width=0.4\textwidth]{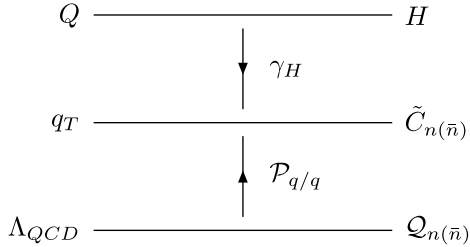}
%\quad\quad\quad
%\includegraphics[width=0.3\textwidth]{dy_large_qt}
\end{center}
\caption{\it
Structure of Drell-Yan factorization theorem. QCD is first matched onto SCET-$q_T$ at the scale $\m\sim Q$ through $H$, followed by the RG running down to scale $\m\sim q_T$, resumming part of the logarithms of $q_T/Q$. Then the TMDPDFs $j_{n(\bn)}$ are matched onto the standard PDFs ${\cal Q}_{n(\bn)}$ at the scale $\m\sim q_T$, in the impact parameter space, through $\tilde C_{n(\bn)}$.
The rest of the logarithms of $q_T/Q$ are resummed by the exponentiation of the $Q^2$-factor.
Finally, the PDFs are evolved from $\m\sim \L_{QCD}$ up to $\m\sim q_T$ via DGLAP equations, resumming logarithms of $\L_{QCD}/q_T$.
\label{dy_small_qt}}
\end{figure}

In the kinematic region where  $\Lambda_{QCD} \ll q_T\ll Q$ the logarithms of the scales ratio need to be resummed to all orders in perturbation theory. For phenomenological applications one also needs to consider the DGLAP evolution of the PDF from a factorization scale up to some intermediate scale $\mu_I$ as illustrated in Fig.~(5). The DGLAP evolution is well-understood and will not be discussed any further below.
In impact parameter space where the factorization theorem becomes a simple product one might be tempted, following the effective field theory methodology, to resum large logarithms of $(q_T^2/Q^2)$
 by evolving the relevant anomalous dimension(s) of the effective theory operator(s). This would be true in the case of threshold resummation however this pattern is not sufficient to resum all logarithms for low-$q_T$ observables. As was pointed out in \cite{Becher:2010tm} and as we mentioned in  the previous subsection, the appearance of the logarithmic $Q^2$-dependent terms in the OPE Wilson coefficients--order by order in perturbation theory--of the TMDPDF onto the integrated PDF complicates the standard EFT resummation procedure since, on one hand, those logarithms do not cancel by any choice of the intermediate scale and, on the other hand, they cannot be resummed by standard RGE equations. They are resummed once they exponentiate and form the ``collinear anomaly'' contribution as illustrated in \cite{Becher:2010tm}.

As we have mentioned in the previous subsection, the origin of such logarithms is attributed, in our case, to the non-vanishing contribution of the soft function to the TMDPDF rather than collinear anomaly. However the resummation of the large logarithms can still be preformed in the same way as is done in \cite{Becher:2010tm}. In both cases we have the hard matching coefficient is identical and also the final form of the factorization theorem (after the OPE is performed.) It is clear then that the resummation procedure for the hadronic tensor can proceed along the same lines. The major difference though, is that in our case it is possible to discuss the resummation of large logarithms contributing to \emph{individual} TMDPDFs rather than to the complete hadronic tensor. This fact is important phenomenologically. One can obtain a resummed TMDPDF in one high-energy process and implement it in a different one due to the universal features of this quantity. More discussion about this is given in Sec.~\ref{sec:universality}.

The resummed hadronic tensor is
\begin{align}\label{eq:mresummed}
M(x,z;\vec q_\perp,Q) &= \int \frac{d^2b_\perp}{(2\pi)^2} e^{-i\vec q_\perp \cdot \vec b_\perp}
\nn\\
&\times
\exp\left[\int_Q^{\m_I} \frac{d\m'}{\m'} \g_H\right]
H(Q^2,\m^2=Q^2)\,
\tilde j_n(x;\vec b_\perp,Q,\m_I)\, \tilde j_\bn(z;\vec b_\perp,Q,\m_I)\,,
\end{align}
where the resummed TMDPDF in impact parameter space is
\begin{align}\label{eq:jnresummed_ips}
\tilde j_n(x;\vec b_\perp,Q,\m) &=
\exp \left[ \int_{\m_I}^{\m} \frac{d\m'}{\m'} \g_n \right]
\tilde j_n(x;\vec b_\perp,Q,\m_I)
\nn\\
&=
\exp \left[ \int_{\m_I}^{\m} \frac{d\m'}{\m'} \g_n \right]
\left( \frac{Q^2 b^2 e^{2\g_E}}{4} \right)^{-D(\as,L_T=0)}
\tilde {\cal C}_n \left(x;\vec b_\perp,\m_I\right) \otimes {\cal Q}_n(x;\m_I)\,.
\end{align}
All the large logarithms in Eq.~(\ref{eq:mresummed}) are contained in the first exponential, the $Q^2$-dependent factor and the evolution of the PDF (in Eq.~(\ref{eq:jnresummed_ips})). When matching QCD onto SCET-$q_T$ we extract the coefficient $H$, and by running it from $Q$ down to $\m_I$ we resum part of the logs of $q_T/Q$. The rest is resummed by the exponentiation of the $Q^2$-dependent factor, which comes from the OPE of the TMDPDF in SCET-$q_T$ onto the PDF in SCET-II. Finally, since we need the PDFs at scale $\m_I$, they are evolved by the standard DGLAP from a lower scale $\L_{QCD}$ up to $\m_I$, resumming all logs of $\L_{QCD}/q_T$. Notice that, due to Eq.~(\ref{eq:ads}), the running of the hard matching coefficient $H$ from $Q$ down to $\m_I$ with $\g_H$, is actually equivalent to the evolution of the two TMDPDFs from $\m_I$ up to $Q$ with $\g_n$ and $\g_{\bn}$.
\footnote{Notice that $\g_{n(\bn)}$ refers to $j_{n(\bn)}$ and not to ${\cal Q}_{n(\bn)}$.}

The AD of the TMDPDF at first order in $\as$ was already given in Eq.~(\ref{g_n}). Based on Eq.~(\ref{eq:ads}), we can extract it from \cite{Idilbi:2005er} at second order in $\as$:
\begin{align}
\g_{n2} &= -\frac{1}{2} \g_{H2} = -\frac{1}{2} 2 \left( \frac{\as}{\pi} \right)^2 \left\{
\left[ \left( \frac{67}{36} - \frac{\pi^2}{12} \right) C_A - \frac{5}{18} N_f \right] C_F \ln\frac{Q^2}{\m^2}
\right.
\nn\\
&\left.
+ \left( \frac{13}{4} \zeta(3) - \frac{961}{16\times 27} - \frac{11}{48}\pi^2 \right) C_A C_F
+ \left( \frac{\pi^2}{24} + \frac{65}{8\times 27} \right) N_f C_F
+ \left( \frac{\pi^2}{4} - \frac{3}{16} - 3\zeta(3) \right) C_F^2
\right\}\,.
\end{align}
The last result and the AD at third order in $\as$, $\g_{n3}$, which can be extracted in the same manner as $\g_{n2}$ from \cite{Moch:2005id} (see also \cite{Idilbi:2006dg,Becher:2006mr}), are essential ingredients to perform phenomenological predictions with higher logarithmic accuracies.
We can write the resummed TMDPDF in momentum space as well,
\begin{align}\label{eq:jnresummed_ms}
j_n(x;\vec k_{n\perp},Q,\m) &=
\int \frac{d^2\vec b_\perp}{(2\pi)^2} e^{-i\vec k_{n\perp} \cdot \vec b_\perp}
\nn\\
&\times
\exp \left[ \int_{\m_I}^{\m} \frac{d\m'}{\m'} \g_n \right]
\left( \frac{Q^2 b^2 e^{2\g_E}}{4} \right)^{-D(\as,L_T=0)}
\tilde {\cal C}_n \left(x;\vec b_\perp,\m_I\right) \otimes {\cal Q}_n(x;\m_I)\,.
\end{align}
Notice that the expression above suffers from the well-known Landau pole when integrating over large values of $b$, since the integrand depends on $\as(\m_I)$. In the literature this issue is generally overcomed by setting a cutoff in $b$ and adding a non-perturbative model function for the contribution from long-distance physics. However, when the resummation is done in momentum space, following the procedure explained in  \cite{Becher:2010tm}, one would expect to sidestep this issue for individual TMDPDF.  The resummation of large logarithms directly in momentum space for individual TMDPDF (and not the product of two) can also be accomplished following similar analysis as the one advocated in \cite{Becher:2010tm} however this issue will be addressed elsewhere.

%%%%%%%%%%%%%%%%%%%%%%%%%%%%%%%%%
%%%%%%%%%%%%%%%%%%%%%%%%%%%%%%%%%
\section{Universality of the TMDPDF}
\label{sec:universality}
%%%%%%%%%%%%%%%%%%%%%%%%%%%%%%%%%
The predictive power of perturbative QCD relies on the universality of the non-perturbative matrix elements that enter the factorization theorems relevant for different high energy processes. Those quantities can be extracted from a limited set of hard reactions and then applied  to make predictions for other processes. In the following we examine the universality properties of the TMDPDFs, Eq.~(\ref{eq:jtmd}), by considering  them in two different kinematical settings: one is for DIS and the other is for DY.

The difference between DIS and DY settings appears already at the level of the operator definitions of the collinear and soft matrix elements of the TMDPDFs due to the existence of different Wilson lines between the two settings. Moreover, and since the soft function connects two collinear sectors, which are obviously different between DIS and DY, it is not immediately clear how the universality of the TMDPDFs is realized.

For completeness we write below the Wilson lines for DIS kinematics, which can be used in actual calculations for the TMDPDF that obeys the DIS setup:
\begin{align}\label{dis_wilson}
&\tilde W_{n(\bn)}^T = \tilde T_{n(\bn)} \tilde W_{n(\bn)}\,,
%\nn\\
%&\tilde S_{\bn}^T (x) = \tilde T_{s\bn} \tilde S_{\bn}\,,
\nn\\
&\tilde W_{n} (x) = \bar P \exp \left[-ig \int_{0}^{\infty} ds\, \nb \cdot A_n (x+\bn s)\right]\,,
%&\quad&
%\tilde W_{n}^\dagger (x) = P \exp \left[-i g \int_{0}^{\infty} ds\, \nb \cdot A_n (x+\bn s)\right]\,,
%\nn\\
%&\tilde W_{\bn} (x) = \bar P \exp \left[ig \int_{0}^{\infty} ds\, n \cdot A_\bn (x+ns)\right]\,,
%&\quad&
%\tilde W_{\bn}^\dagger (x) = P \exp \left[-ig \int_{0}^{\infty} ds\, n \cdot A_\bn (x+ns)\right]\,,
\nn\\
&\tilde T_{n} (x) = \bar P \exp \left[-ig \int_{0}^{\infty} d\tau\, \vec l_\perp \cdot \vec A_{n\perp} (x^+,\infty^-,\vec x_\perp+\vec l_\perp \tau)\right]\,,
%&\quad&
%\tilde T_{n}^\dagger (x) = P \exp \left[-ig \int_{0}^{\infty} d\tau l_\perp \cdot A_{n\perp} (x^+,\infty^-,x_\perp+l_\perp \tau)\right]\,,
\nn\\
%&\tilde T_{\bn} (x) = \bar P \exp \left[ig \int_{0}^{\infty} d\tau l_\perp \cdot A_{\bn\perp} (\infty^+,x^-,x_\perp+l_\perp \tau)\right]\,,
%&\quad&
%\tilde T_{\bn}^\dagger (x) = P \exp \left[-ig \int_{0}^{\infty} d\tau l_\perp \cdot A_{\bn\perp} (\infty^+,0^-,x_\perp+l_\perp \tau)\right]\,,
%\nn\\
&\tilde S_\bn (x) = P\exp\le[-ig\int_{0}^{\infty} ds\, \bn \cdot A_s(x+\bn s) \ri]\,,
%&\quad&
%\tilde S_\bn^\dagger (x) = \bar P \exp \le[ig\int_{0}^{\infty} ds\, \bn \cdot A_s(x+\bn s) \ri]\,,
\nn\\
&\tilde T_{s\bn} (x) = P\exp\le[-ig\int_{0}^{\infty} d\t\, \vec l_\perp \cdot \vec A_{s\perp}(0^+,\infty^-,\vec x_\perp+\vec l_\perp\t) \ri]\,,
%&\quad&
%\tilde T_{s\bn}^\dagger (x) = \bar P\exp\le[ig\int_{0}^{\infty} d\t\, l_\perp \cdot A_{s\perp}(x+l_\perp\t) \ri]\,.
\end{align}
where the rest of the Wilson lines can be obtained by exchanging $n \leftrightarrow \bn$ and $P\leftrightarrow \bP$ and the relevant matrix elements for the TMDPDF in DIS are
\begin{align}\label{dis_elements}
\hat f_n^{DIS} &= \langle p|\,\le[\bar\x_n \tilde W^T_n\ri](0^+,r^-,\vec r_\perp)\,\frac{\bnslash}{2}\,
\le[\tilde W_n^{T\dagger} \x_n\ri](0)\,|p\rangle|_\textrm{zb~ included}\,,
\nn\\
\hat f_\bn^{DIS} &= \langle \bar p|\,\le[\bar\x_\bn \tilde W^T_\bn\ri](0)\,\frac{\nslash}{2}\,
\le[\tilde W_\bn^{T\dagger} \x_\bn\ri](r^+,0^-,\vec r_\perp)\,
|\bar p\rangle|_\textrm{zb~ included}\,,
\nn\\
\phi_{DIS} &= \sandwich{0}{{\rm Tr}\, \le[\tilde S_\bn^T S_n^{T\dagger}\ri](0^+,0^-,\vec r_\perp)\le[\tilde S^{T\dagger}_\bn S_n^T\ri](0)}{0}\,.
\end{align}
Notice that all the collinear Wilson lines are different between DY and DIS, unlike the soft ones, where only the $\bn$-soft Wilson line changes. This is due to the fact that
collinear Wilson line, say $W_n$, ``knows'' about the collinearity of the $\bn$-sector, while the soft ones are related just to their own sector.

In this section we will show to first order in $\as$ that the TMDPDF, defined in Eqs.~(\ref{eq:jtmdgeneral}) and Eq.~(\ref{eq:jtmd}), is the same in DY and DIS kinematics. This expectation is based on the results of Sec.~\ref{sec:tmdpdftopdf}. In that section we have shown that when we integrate the TMDPDF over $k_{n\perp}$ we recover the standard PDF. It is well known that the PDF is universal. Thus, it is expected that our TMDPDF will also be universal (at least to first order in $\as$.) Below we show this to hold using the $\d$-regulator and in momentum space.

The virtual contribution to the TMDPDF in the case of DY is given in Eq.~(\ref{eq:jvn}). The real contribution comes from the combination of Eqs.~(\ref{eq:jn3a}, \ref{eq:jn3b3c}, \ref{eq:jn4b4c}) according to Eq.~(\ref{jn}).

The TMDPDF that enters the factorization theorem for DIS at low $q_T$ spectrum is
\begin{align}
j_n^{DIS}(x;\vec k_{n\perp}) &=\frac{1}{2}
\int \frac{dr^-d^2\vec r_\perp}{(2\pi)^3} e^{-i(\frac{1}{2}r^-xp^+-\vec r_\perp \cdot \vec k_{n\perp})}
\frac{\hat f_n^{DIS}(0^+,r^-,\vec r_\perp)}{\sqrt{\phi^{DIS}(0^+,0^-,\vec r_\perp)}}\,,
\end{align}
where $\hat f^{DIS}_n$ and $\phi^{DIS}$ were defined above.
The diagrams in figs.~(\ref{n_virtuals}) and~(\ref{s_virtuals}) give collinear and soft virtual contributions respectively to $j_n^{DIS}$. The WFR diagram~(\ref{n_virtuals}a) and its Hermitian conjugate give
\begin{align}
\hat f_{n1}^{DIS,(\ref{n_virtuals}a)}&=
\frac{\alpha_s C_F}{2\pi}
\d(1-x)\d^{(2)}(\vec k_{n\perp})
\le[ \frac{1}{2\veuv}+\frac{1}{2}\ln\frac{\m^2}{-i\D^-}+\frac{1}{4} \ri] + h.c.
\end{align}
The $W$ Wilson line tadpole diagram, (\ref{n_virtuals}b), is identically $0$, since $\bn^2=0$.
Diagram~(\ref{n_virtuals}c)  and its Hermitian conjugate give
\begin{align}
\hat f_{n1}^{DIS,(\ref{n_virtuals}c)}&=
-2ig^2C_F \d(1-x)\d^{(2)}(\vec k_{n\perp}) \m^{2\e} \int \frac{d^dk}{(2\pi)^d}
\frac{p^++k^+}{[k^++i\d^+][(p+k)^2+i\D^-][k^2+i0]}
+ h.c.\nn\\
&=
\frac{\a_s C_F}{2\pi}
\d(1-x)\d^{(2)}(\vec k_{n\perp})
\left[
\frac{2}{\veuv}\ln\frac{\d^+}{p^+} + \frac{2}{\veuv} - \ln^2\frac{\d^+\D^-}{p^+\m^2}
- 2\ln\frac{\D^-}{\m^2} + \ln^2\frac{\D^-}{\m^2} + 2 + \frac{5\pi^2}{12}
\right]\, .
\end{align}
The contribution of diagrams~(\ref{s_virtuals}a) and~(\ref{s_virtuals}b) is zero, since~(\ref{s_virtuals}a) is proportional to $n^2=0$
and~(\ref{s_virtuals}b) to $\bn^2=0$.
The diagram~(\ref{s_virtuals}c) and its Hermitian conjugate give
\begin{align}
\phi_1^{DIS,(\ref{s_virtuals}c)}&=
-2ig^2 C_F \d^{(2)}(\vec k_{n\perp}) \mu^{2 \eps}
\int \frac{d^d k}{(2 \pi)^d} \frac{1}{[k^++i\d^+] [k^-+i\d^-] [k^2+i0]} +h.c.
\nn \\
&=
- \frac{\alpha_s C_F}{2\pi}
\d^{(2)}(\vec k_{n\perp})
\left[\frac{2}{\veuv^2}-\frac{2}{\veuv}\ln\frac{\d^+\d^-}{\mu^2}+
\ln^2\frac{\d^+\d^-}{\mu^2}-\frac{\pi^2}{2}\right]\, .
\end{align}
Thus, the results for the virtual contribution for naive collinear, soft, TMDPDF and pure collinear are, respectively,
\begin{align}
\label{eq:resultsdis}
\hat f_{n1}^{v,DIS} &=
\hat f_{n1}^{DIS,(\ref{n_virtuals}c)} - \frac{1}{2} \hat f_{n1}^{DIS,(\ref{n_virtuals}a)}
\nn\\
&=
\frac{\a_s C_F}{2\pi}
\d(1-x)\d^{(2)}(\vec k_{n\perp})
\left[
\frac{2}{\veuv}\ln\frac{\D}{Q^2} + \frac{3}{2}\frac{1}{\veuv} - \ln^2\frac{\D^2}{Q^2\m^2}
- \frac{3}{2}\ln\frac{\D}{\m^2} + \ln^2\frac{\D}{\m^2}+ \frac{7}{4} + \frac{5\pi^2}{12}
\right]\,,
\nn\\
\phi_{1}^{v,DIS} &= \phi_1^{DIS,(\ref{s_virtuals}c)}
=
- \frac{\a_sC_F}{2\pi}
\d^{(2)}(\vec k_{n\perp})
\le[
\frac{2}{\veuv^2} - \frac{2}{\veuv}\ln\frac{\D^2}{Q^2\m^2} + \ln^2\frac{\D^2}{Q^2\m^2} - \frac{\pi^2}{2} \ri]\,,
\nn\\
j_{n1}^{v,DIS} &=
\frac{\alpha_s C_F}{2 \pi}
\d(1-x)\d^{(2)}(\vec k_{n\perp})
\left[
\frac{1}{\veuv^2} + \frac{1}{\veuv} \left( \frac{3}{2} + \ln\frac{\mu^2}{Q^2} \right)
- \frac{3}{2}\ln\frac{\D}{\mu^2} - \frac{1}{2}\ln^2\frac{\D^2}{Q^2\m^2} + \ln^2\frac{\D}{\m^2}
+ \frac{7}{4} + \frac{\pi^2}{6}\right]\,,
\nn\\
f_{n1}^{v,DIS} &=
\frac{\a_s C_F}{2\pi}
\d(1-x)\d^{(2)}(\vec k_{n\perp})
\le[ \frac{2}{\veuv^2} + \frac{1}{\veuv} \le( \frac{3}{2} + 2\ln\frac{\m^2}{\D} \ri) - \frac{3}{2}\ln\frac{\D}{\m^2}
+ \ln^2\frac{\D}{\m^2} + \frac{7}{4} - \frac{\pi^2}{12} \ri]\,.
\nn\\
\end{align}
Notice that the pure collinear matrix element given in the last line above is calculated by subtracting the complete soft function from the naive collinear matrix element. For DY one can get the same result.
Comparing our results for DY kinematics with DIS ones we get the following
\begin{align}
\label{eq:relnaive}
\hat f_{n1}^{v,DIS} &=
\hat f_{n1}^{v} + \frac{\as C_F}{2\pi}\d(1-x)\d^{(2)}(\vec k_{n\perp})\, \pi^2\,,
\end{align}
and
\begin{align}
\label{eq:relsoft}
\phi_{1}^{v,DIS} &=
\phi_{1}^{v} + \frac{\as C_F}{2\pi}\d^{(2)}(\vec k_{n\perp})\, \pi^2\,.
\end{align}

We show below that the real part of the naive collinear and soft matrix elements are also different, and that this difference exactly compensates the one in the virtual parts. We remind the reader that all the results below are valid for infinitesimally small  $\D^\pm$ with respect to all other scales.
Diagram (\ref{n_reals}a) is the same for DY and DIS,
\begin{align}
\hat f_{n1}^{(\ref{n_reals}a)} &= \hat f_{n1}^{DIS,(\ref{n_reals}a)} =
\frac{\a_s C_F}{2\pi^2}
(1-\ve)(1-x)
\frac{|\vec k_{n\perp}|^2}{\left||\vec k_{n\perp}|^2-i\D^-(1-x)\right|^2}\,.
\end{align}
The contribution of diagrams~(\ref{n_reals}b+\ref{n_reals}c) for DY was given before and it can be expressed as
\begin{align}
\hat f_{n1}^{(\ref{n_reals}b+\ref{n_reals}c)}&=
\frac{\as C_F}{2\pi^2}
\left[\frac{x}{(1-x)+i\d^+/p^+}\right]
\left[
\frac{1}{|\vec k_{n\perp}|^2-i\D^-(1-x)}
\right] + h.c.
\nn\\
&=
\frac{\as C_F}{2\pi^2}
\left\{
PV\left( \frac{1}{|\vec k_{n\perp}|^2}\right)
\left[ \frac{x}{(1-x)+i\d^+/p^+}+\frac{x}{(1-x)-i\d^+/p^+} \right]
\right.
\nn\\
&\left.
+ i\pi\d(|\vec k_{n\perp}|^2) \big( -i\pi\d(1-x) \big)\right\}
\,,
\end{align}
while for DIS it is
\begin{align}
\hat f_{n1}^{DIS,(\ref{n_reals}b+\ref{n_reals}c)}&=
-4\pi g^2 C_F p^+  \int\frac{d^dk}{(2\pi)^d}
\d(k^2)\theta(k^+)\frac{p^+-k^+}{[k^+-i\d^+][(p-k)^2+i\D^-]}
\nn\\
&\times
\d\le((1-x)p^+-k^+\ri) \d^{(2)}(\vec k_\perp+\vec k_{n\perp}) + h.c.
\nn\\
&=
\frac{\as C_F}{2\pi^2}
\left[\frac{x}{(1-x)-i\d^+/p^+}\right]
\left[
\frac{1}{|\vec k_{n\perp}|^2-i\D^-(1-x)}
\right] + h.c.
\nn\\
&=
\frac{\as C_F}{2\pi^2}
\left\{
PV\left( \frac{1}{|\vec k_{n\perp}|^2}\right)
\left[ \frac{x}{(1-x)+i\d^+/p^+}+\frac{x}{(1-x)-i\d^+/p^+} \right]
\right.
\nn\\
&\left.
+ i\pi\d(|\vec k_{n\perp}|^2) \left( +i\pi\d(1-x) \right)\right\}
\,.
\end{align}
Thus, the real part of the naive collinear matrix elements in DY and DIS kinematics are related by the following
\begin{align}
\label{eq:relrealcol}
\hat f_{n1}^{r,DIS} &=
\hat f_{n1}^{r} - \frac{\as C_F}{2\pi^2} \d(1-x)\d(|\vec k_{n\perp}|^2)2\pi^2
\nn\\
&=
\hat f_{n1}^{r} - \frac{\alpha_s C_F}{2\pi}\d(1-x) \d^{(2)}(\vec k_{n\perp})\, \pi^2
\,,
\end{align}
where we have used: $\d(|\vec k_{n\perp}|^2) =(\pi/2) \d^{(2)}(\vec k_{n\perp})$.
Combining the last result with Eq.~(\ref{eq:relnaive}) we conclude that the naive collinear matrix element is universal to ${\cal O}(\as)$.

The soft contribution in diagrams~(\ref{s_reals}b+\ref{s_reals}c) for DY was given in Eq.~(\ref{eq:jn4b4c})
%\begin{align}
%\label{eq:softrealdy}
%\phi_1^{(\ref{s_reals}b+\ref{s_reals}c)}&=
%-\frac{\a_s C_F}{\pi^2}
%\frac{1}{|\vec k_{n\perp}|^2-\d^+\d^-} \ln\frac{\d^+\d^-}{|\vec k_{n\perp}|^2} \,,
%\end{align}
while for DIS we have
\begin{align}
\label{eq:softrealdis}
\phi_1^{DIS,(\ref{s_reals}b+\ref{s_reals}c)}&=
-4\pi g^2 C_F  \int\frac{d^dk}{(2\pi)^d}
\d^{(2)}(\vec k_\perp+\vec k_{n\perp}) \d(k^2)\theta(k^+) \frac{1}{[k^+-i\d^+][-k^-+i\d^-]} + h.c.
\nn\\
&=
-\frac{\a_s C_F}{\pi^2}
\frac{1}{|\vec k_{n\perp}|^2+\d^+\d^-} \ln\frac{\d^+\d^-}{|\vec k_{n\perp}|^2}\,.
\end{align}
Since we are interested in expressing our results in terms of distributions in momentum space, it turns out to be easier to consider the difference between the real contribution of the soft function for DY and DIS.
In order to achieve this let us write the following:
\begin{align}
\phi_1^{(\ref{s_reals}b+\ref{s_reals}c)} &=
-\frac{\a_s C_F}{\pi^2} g^{DY}(a;t)\,, \quad\quad\quad
g^{DY}(a;t) = -\frac{1}{a}\frac{\ln t}{t-1}\,,
\nn\\
\phi_1^{DIS,(\ref{s_reals}b+\ref{s_reals}c)} &=
-\frac{\a_s C_F}{\pi^2} g^{DIS}(a;t)\,, \quad\quad\quad
g^{DIS}(a;t) = -\frac{1}{a}\frac{\ln t}{t+1}\,,
\end{align}
where $t=|\vec k_{n\perp}|^2/(\d^+\d^-)$ and $a=\d^+\d^-$.
One can easily see that
\begin{align}
g^{DIS}(a;t) - g^{DY}(a;t) = A\, \d^{(2)}(\vec k_{n\perp})\,,
\end{align}
and integrating over $\vec k_{n\perp}$ we get the coefficient A,
\begin{align}
A &=
\int d^2k_{n\perp} \left[ g^{DIS}(a;t) - g^{DY}(a;t) \right] =
a\pi \int_0^{\infty} dt \left[ g^{DIS}(a;t) - g^{DY}(a;t) \right] =
\nn\\
&=
2\pi \int_0^{\infty} dt \frac{\ln t}{t^2-1}
=
\frac{\pi^3}{2}\,.
\end{align}
The functions $g^{DY}$ and $g^{DIS}$ are UV-divergent when integrated over $\vec k_{n\perp}$. However when we take the difference, we get a UV-finite contribution and IR-regularized with the $a$ parameter. This is due to the fact that the difference between DY and DIS integrals is just the position of the pole of the collinear Wilson line, which is related to the IR (collinear) divergence.
Thus, the real contributions to the soft functions for DY and DIS kinematics are related according to
\begin{align}
\label{eq:relrealsoft}
\phi_1^{r,DIS} &= \phi_1^{r}
- \frac{\a_s C_F}{2\pi}\d^{(2)}(\vec k_{n\perp})\, \pi^2\,.
\end{align}
Combining this result with Eq.~(\ref{eq:relsoft}) we conclude that the soft function is universal to ${\cal O}(\as)$.

To conclude this section, we have shown that the naive collinear and the soft are universal, from which the pure collinear and the TMDPDF are clearly universal. In the Appendix we calculate the TMDPDF in impact parameter space for DY and DIS, and then match it onto the PDF, where all those quantitates are calculated with the $\d$-regulator. By doing so, we show that the PDF is universal, as it should be, and that the matching coefficient at the intermediate scale is the same for DY and DIS kinematics and independent of the IR regulator.

%%%%%%%%%%%%%%%%%%%%%%%%%%%%%%%%
%%%%%%%%%%%%%%%%%%%%%%%%%%%%%%%%
\section{Factorization at ${\cal O}(\as)$}
\label{sec:factorization}
%%%%%%%%%%%%%%%%%%%%%%%%%%%%%%%%

In this Section we establish the factorization theorem given in Eq.~(\ref{eq:mainfact}) to first order in $\as$. We do it through Eq.~(\ref{step42}), since we have already established the OPE of the TMDPDFs in Eq.~(\ref{eq:ope}) given the results in Eqs.~(\ref{eq:pdfir},\ref{coeff}). The hard matching coefficient for the $q_T$-dependent DY cross section is the same as the one for inclusive DY. As mentioned before, this matching coefficient at the higher scale $Q$ is obtained by matching the full QCD cross section onto the imaginary part of the product of two effective theory currents. This echoes the ``subtraction method'' in perturbative QCD.

We start by rewriting Eq.~(\ref{step42}) in a more useful way,
\begin{align} \label{eq:hadten}
d\sigma&=\frac{4 \pi\alpha}{3 N_c q^2 s}\frac{dx dz d^2 \vec q_\perp}{2 (2\pi)^4}
\sum_q e_q^2 M(x,z;\vec q_\perp,Q)\,,
\nn\\
M(x,z;\vec q_\perp,Q)&=
H(Q^2/\mu^2)
\int d^2\vec k_{n\perp} d^2\vec k_{\bn\perp}\,
\d^{(2)}(\vec q_\perp-\vec k_{n\perp}-\vec k_{\bn\perp})
\left[
%j_{n0} j_{\bn 0}
 \d(1-x)\d^{(2)}(\vec k_{n\perp})\d(1-z)\d^{(2)}(\vec k_{\bn\perp})
\right. \nn \\
&\left.
 + \a_s \left( j_{n1}\,\d(1-z)\d^{(2)}(\vec k_{\bn\perp}) +
  j_{\bn 1}\, \d(1-x)\d^{(2)}(\vec k_{n\perp}) \right)
\right] +O(\a_s^2)\,
\nn \\
&=H(Q^2/\mu^2)
\left[
\d(1-x)\d(1-z)\d^{(2)}(\vec q_\perp) \right.
\nn\\
&\left.
+
\a_s \Big(
\d(1-z)\, j_{n1}(x;\vec q_\perp,Q,\m) +
\d(1-x)\, j_{\bn 1}(z;\vec q_\perp,Q,\m)
\Big)
\right]
+O(\a_s^2)\,,
\end{align}
where $M$ is the hadronic tensor.

The real parts of $j_n$ and $j_\bn$ at order $\a_s$ are given in Eqs.~(\ref{eq:jr})-(\ref{eq:jrnb}) and using Eq.~(\ref{eq:hadten}) one obtains the total real part of the hadronic tensor $M$ in the effective theory,
\begin{align}
M_{SCET}^r&=
\frac{\a_s C_F}{2\pi^2} \frac{1}{q_T^2} \left[
\d(1-x)(1-z) + \d(1-z)(1-x) + \d(1-x)\frac{2z}{(1-z)_+}
\right.
\nn\\
&\left.
+ \d(1-z)\frac{2x}{(1-x)_+}
+ 2\d(1-x)\d(1-z)\ln\frac{Q^2}{q_T^2}
\right]\, .
\end{align}
The real part contribution for DY at non-vanishing $q_T$ in QCD can be read--off from the result given in Ref.~\cite{Ji:2004wu} (see Eqs.~(57)-(59) in that reference) by going from space--like (DIS) to time--like (DY) kinematics. Since the QCD result for DIS includes only single log, then the results for $M^r$ at ${\cal O}(\as)$ are equal in both cases. Thus, one can easily see that $M^r_{SCET} = M^r_{QCD}$.

In QCD the virtual part of $M$ with the $\delta$-regulator is
\begin{align}\label{eq:mqcdv}
M_{QCD}^v =
\frac{\a_s C_F }{2\pi}
\d(1-x)\d(1-z)\d^{(2)}(\vec q_\perp)
\left[ - 2\ln^2\frac{\D}{Q^2} - 3\ln\frac{\D}{Q^2} - \frac{9}{2} + \frac{\pi^2}{2} \right]\,.
\end{align}
The above result can be simply obtained by considering the one-loop correction to the vertex diagram for $q\bar q \to \g^*$, with the inclusion of the WFR diagram while using the fermion propagators in Eq.~(\ref{fermionsDelta}) with $\D^\pm=\D$.

The virtual parts of $j_n$ and $j_\bn$ at one-loop are given in Eqs.~(\ref{eq:jvn}, \ref{eq:jvnb}). Using Eq.~(\ref{eq:hadten}) the total virtual part of the hadronic tensor $M$ in the effective theory is
\begin{align}\label{eq:mscetv}
M_{SCET}^v &=
H(Q^2/\mu^2)
\frac{\a_s C_F}{2\pi}
\d(1-x)\d(1-z)\d^{(2)}(\vec q_\perp)
\left[
\frac{2}{\veuv^2} + \frac{1}{\veuv} \left( 3 + 2\ln\frac{\m^2}{Q^2}  \right) \right.
\nn\\
&
\left.
- 2\ln^2\frac{\D}{Q^2} - 3\ln\frac{\D}{Q^2}
+ 3\ln\frac{\m^2}{Q^2} + \ln^2\frac{\m^2}{Q^2} + \frac{7}{2} - \frac{2\pi^2}{3}
\right]\,,
\end{align}
where the UV divergences are canceled by the standard renormalization process. We notice that the IR contributions in Eqs.~(\ref{eq:mqcdv}, \ref{eq:mscetv}) are the same, thus the matching coefficient between QCD and the effective theory at scale $Q$ is:
\begin{align}
H(Q^2/\mu^2) =
1 + \frac{\a_s C_F}{2\pi} \left[
- 3\ln\frac{\m^2}{Q^2} - \ln^2\frac{\m^2}{Q^2} - 8 + \frac{7\pi^2}{6}
\right]\,.
\end{align}
The above result was first derived in~\cite{Idilbi:2005ky,hep-ph/0309278}. We can also obtain the AD of the hard matching coefficient at ${\cal O}(\as)$ and verify Eq.~(\ref{eq:ads}),
\begin{align}
\g_{H1} = - \frac{\as C_F}{2\pi} \left[ 6 + 4\ln\frac{\m^2}{Q^2} \right] = -2\g_{n1}\,.
\end{align}

So we conclude that the factorization theorem in Eq.~(\ref{eq:mainfact}) is satisfied to first order in $\as$. The IR divergences of full QCD are recovered in the effective theory calculation, Eq.~(\ref{eq:mscetv}). The real contribution is the same in QCD and in the effective theory, and finally, the matching coefficient at the higher scale depends only on the hard scale $Q^2$ as it should be.

%%%%%%%%%%%%%%%%%%%%%%%%%%%%%%%
%%%%%%%%%%%%%%%%%%%%%%%%%%%%%%%
\section{Light-Cone Gauge}
\label{sec:lcg}
%%%%%%%%%%%%%%%%%%%%%%%%%%%%%%%
\begin{figure}
\begin{center}
\includegraphics[width=0.7\textwidth]{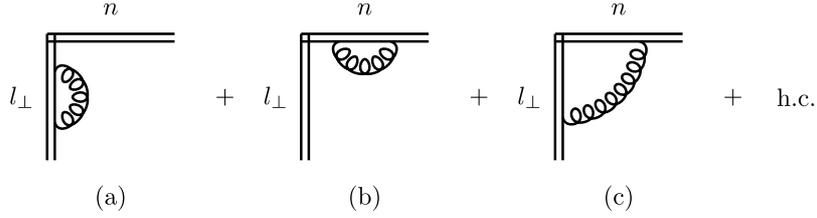}
\end{center}
\caption{\it The  soft  function at one-loop in light-cone gauge.
\label{s_virtuals_lc}}
\end{figure}

 %In Ref.~\cite{Cherednikov:2007tw} the authors pretend that the light-cone gauge has a special role in  the  demonstration
%of the factorization theorems.  We disagree with this result, but before commenting on the results of these authors

In this Section we show that the TMDPDFs, $J_{n(\bn)}$, are actually  the same in light-cone  gauge and Feynman gauge, once the contribution from the transverse Wilson lines is taken into account. In Ref.~\cite{Idilbi:2010im} two of us have shown that the naive collinear contribution to the TMDPDF (the numerator in $j_{n(\bn)}$) is actually  gauge invariant
 with a one-loop calculation.  In that article  the authors used a particular IR regulator for light-cone divergences
however the results obtained in covariant gauge and in light-cone gauge are the same  and independent of that regulator once the zero-bin corrections are included. That was shown explicitly
in the Appendix of that work.

 In light-cone gauge we use the ML prescription~\cite{Mandelstam:1982cb}, which is the only one consistent
 with the canonical quantization of  QCD in this gauge~\cite{Bassetto:1984dq}.
 Moreover in the $n$ and $\bn$ collinear sectors the only gauge fixings compatible with
the power counting of the collinear particles are respectively $\bn A_n=0$ and $n A_\bn=0$,
which correspond to ``killing'' the highly oscillating component of the gluon field in each sector.
We now compare  the integrals  that we have evaluated in Feynman gauge with the  corresponding ones in light-cone gauge.

The interesting contribution to the collinear part of the TMDPDF in Feynman gauge is provided  by
the $W$ Wilson line and it is  (cfr.~Eq.~(\ref{eq:jn1c}))
\begin{align}
\label{eq:vFey}
\hat f_{n1}^{(\ref{n_virtuals}c)\; (Feyn)}=
-\d(1-x)\d^{(2)}(\vec k_{n\perp}) 2i g^2 C_F \mu^{2\eps}
\int  \frac{d^d k}{(2\pi)^d}\; \frac{1}{(k^2+i0)( k^+-i0)}\frac{ p^++ k^+}{(p+k)^2+i0}\ .
\end{align}
In light-cone-gauge this result  is reproduced  when combining  the axial part of the WFR,
\begin{align}
\label{eq:IAx}
\hat f_{n1}^{(\ref{n_virtuals}a)\; (Ax)}=
\d(1-x)\d^{(2)}(\vec k_{n\perp}) 4 ig^2 C_F \mu^{2\eps}
\int \frac{d^d k}{(2\pi)^d}\; \frac{1}{(k^2+i0)}\frac{ p^++ k^+}{(p+k)^2+i0}\Big[\frac{\theta(k^-)}{k^++i0}+\frac{\theta(-k^-)}{k^+-i0}\Big]\ ,
\end{align}
and the contribution of the $T$ Wilson line is
\begin{align}
\label{eq:IT}
\hat f_{n1}^{(\ref{n_virtuals}c)\; (T)}=
- \d(1-x)\d^{(2)}(\vec k_{n\perp}) 2 ig^2 C_F \mu^{2\eps}
\int \frac{d^d k}{(2\pi)^d}\; \frac{1}{(k^2+i0)}\frac{ p^++ k^+}{(p+k)^2+i0}\theta(k^-)\Big[\frac{1}{k^+-i0}-\frac{1}{k^++i0}\Big]\ .
 \end{align}
It is evident that $\hat f_{n1}^{(\ref{n_virtuals}c)\; (Feyn)}=\hat f_{n1}^{(\ref{n_virtuals}c)\; (T)}-\hat f_{n1}^{(\ref{n_virtuals}a)\; (Ax)}/2$.
The tadpole diagram is null also in light cone gauge since the gluon field does not propagate at infinity~\cite{Idilbi:2010im}.

In Ref.~\cite{GarciaEchevarria:2011md} we have shown that  we need the $T$ Wilson lines also in the soft sector.
We show this  explicitly  by considering the virtual corrections to the soft function using the gauge fixing $\bn A_s=0$.
The  only one-loop virtual correction in Feynman gauge comes from fig.~(\ref{s_virtuals}c)
\begin{align}
\phi_1^{(\ref{s_virtuals}c)}&=
- \d^{(2)}(\vec k_{n\perp}) 2 i g^2 C_F \mu^{2 \eps}\! \int \! \frac{d^d k}{(2 \pi)^d} \frac{1}{ k^-+i 0}
\frac{1}{ k^+-i 0}
\frac{1}{k^2+i 0} \, .
\end{align}
In light-cone gauge we have two types of  contributions: one from the tadpole diagram in fig.~(\ref{s_virtuals_lc})
and the other one is from the $T$ Wilson line.
The  tadpole contribution  in fig.~(\ref{s_virtuals_lc}b) is zero because the transverse gluon fields do not propagate at
infinity as mentioned earlier.
Explicitly, the tadpole contribution from fig.~(\ref{s_virtuals_lc}a) is
\begin{align}
 \phi_1^{(\ref{s_virtuals_lc}a)}&=
 - \d^{(2)}(\vec k_{n\perp}) 2 i g^2 C_F \mu^{2 \eps} \int \frac{d^d k}{(2 \pi)^d} \frac{1}{[k^+]_{ML}}
\frac{1}{  k^-+i  0}\frac{ k^-}{ k^--i  0}
\frac{1}{k^2+i 0}\,
\nn \\
&=
- \d^{(2)}(\vec k_{n\perp}) 2 i g^2 C_F \mu^{2 \eps}\int \frac{d^d k}{(2 \pi)^d} \Big(
\frac{\theta (k^-)}{ k^++i 0}+\frac{\theta (- k^-)}{k^+-i 0}\Big) %\nn \\ &\times
\frac{1}{  k^-+i  0}\frac{ k^-}{  k^--i  0}
\frac{1}{k^2+i 0}
\nn \\
&= 0  \, ,
\end{align}
because when integrating over $k^+$ all poles lie on the same side of the complex plane.
Finally the contribution of the $T$ Wilson line in fig.~(\ref{s_virtuals_lc}c) is
\begin{align}
\phi_1^{(\ref{s_virtuals_lc}c)}&=
- \d^{(2)}(\vec k_{n\perp}) 2 i g^2 C_F \mu^{2 \eps}\int \frac{d^d k}{(2 \pi)^d} \Big(
\frac{\theta (k^-)}{ k^+-i 0}-\frac{\theta (k^-)}{ k^++i 0}\Big)%\nn \\ &\times
\frac{1}{  k^-+i  0}
\frac{1}{k^2+i 0} \ .
\end{align}
Notice that we can add to $\phi_1^{(\ref{s_virtuals_lc}c)}$ the quantity $I^T \equiv 0$ which is defined as
\begin{align}
I^T&= - \d^{(2)}(\vec k_{n\perp}) 2ig^2 C_F \mu^{2 \eps}\int \frac{d^d k}{(2 \pi)^d} \Big(
\frac{\theta (  k^-)}{ k^++i 0}+\frac{\theta (-  k^-)}{k^+-i 0}\Big)% \nn \\ &\times
\frac{1}{k^-+i  0}
\frac{1}{k^2+i 0}  \, .
\end{align}
 The quantity $I^T$ is exactly zero because when integrating in $ k^+$ all poles, again, lie on the same side of the complex plane.
 Now it is easy to verify that $\phi_1^{(\ref{s_virtuals}c)}=\phi_1^{(\ref{s_virtuals_lc}c)}+I^T$ at the level of integrands.
 In other words  the $T$ Wilson lines in the soft  sector insure the  gauge invariance of  the soft matrix element
irrespective of any infrared regulator.
Similar considerations hold for Feynman diagrams with real gluon contributions.

As a final comment let us consider the work of Ref.~\cite{Cherednikov:2007tw}
 where the authors consider the definition of TMDPDF  in light-cone gauge. In that work it is argued that in order to properly define the TMDPDF one should divide the collinear matrix element by the soft function (and not by the square root of it.)
  However in this Section we have shown  explicitly that the light-cone gauge provides a reshuffling of Feynman diagrams,
while the final results for matrix elements are the same in all gauges. We have also shown that when subtracting the square root of the soft function, as defined in this work, one can achieve a separation of UV and light-cone divergences. It is not clear to us how this feature is realized in  Ref.~\cite{Cherednikov:2007tw}.

%%%%%%%%%%%%%%%%%%%%%%%%%%%%%%%%%
\section{Equivalence of Soft and Zero-Bin Subtractions}
\label{sec:zerobin}
%%%%%%%%%%%%%%%%%%%%%%%%%%%%%%%%%

Let us start this discussion by considering Eq.~(\ref{eq:jn3b3c}) which gives the non-trivial real gluon emission to the naive collinear contribution to the TMDPDF. When taking the gluon momentum $k$ to the soft limit: $k\sim Q(\lambda,\lambda,\lambda)$, one needs then to distinguish between generic values  $1-x$ where it scales as $1$, on one hand, and the threshold region where $1-x$ scales as $\lambda$ on the other. In the former case, taking the soft or zero-bin limit amounts to dropping the $k^+$ from the $\delta((1-x)p^+-k^+)$ thus getting a trivial $\delta(1-x)$ contribution. In this case the equivalence of soft and zero-bin subtractions can be easily verified, as we show below. However at the threshold region and in the soft limit the term $\delta((1-x)p^+-k^+)$ remains intact. This will give a non-trivial $x$-dependence, manifested not only by $\delta(1-x)$ but also with the appearance of $1/(1-x)_+$ in the zero-bin contribution at ${\cal{O}}(\alpha_s)$ and with more involved ``$+$'' distributions at higher orders. Given that our soft function is independent of $x$, then the equivalence of soft and zero-bin subtractions breaks down. This is in complete contrast to the case of partonic observables at threshold. In the latter, the soft function has to have an explicit $x$-dependence--which arises from separation of the soft Wilson lines in the soft function along one light-cone direction--and this dependence is fundamental to establish the equivalence of soft and zero-bin subtractions \cite{Idilbi:2007ff}.

Moreover, when certain IR regulators are implemented different results for the soft and zero-bin contributions are obtained. In \cite{Chiu:2012ir} the zero-bin is zero beyond tree-level while the soft function has non-vanishing contributions to all orders in perturbation theory. Below we establish the equivalence of the zero-bin and the soft function subtractions at order $\a_s$ while staying on-the-light-cone and using the $\d$-regulator. The key point to notice is the relation between the regulators in both collinear sectors, Eq.~(\ref{regul_DeltaDY}).

The pure collinear matrix element $f_n$ is calculated by first integrating over all momentum space and then subtracting the soft limit. Clearly this is done on a diagram-by-diagram basis and perturbatively,
\begin{align}
f_n = \hat f_{n0} + \le( \hat f_{n1} - \hat f_{n1,zb} \ri) + {\cal O}(\alpha_s^2)\,.
\end{align}
We show below to $O(\a_s)$ that this can be achieved by dividing the naive collinear matrix element by the soft function
\begin{align}
f_n = \frac{\hat f_n}{\phi} = \hat f_{n0}  + \le( \hat f_{n1} - \hat f_{n0}\,\phi_1 \ri) + {\cal O}(\alpha_s^2)\,.
\end{align}
For $f_\bn$ analogous analysis trivially applies. The zero-bin of the WFR in diagram~(\ref{n_virtuals}a) is zero, and also the one for diagram~(\ref{n_virtuals}b). The zero-bin of diagram~(\ref{n_virtuals}c) is obtained by setting the loop momentum to be soft, $k\sim(\l,\l,\l)$
\begin{align}
\hat f_{n1,zb}^{(\ref{n_virtuals}c)} &=
- \d(1-x)\d^{(2)}(\vec k_{n\perp}) 2ig^2C_F \m^{2\e}\int\frac{d^dk}{(2\pi)^d}
\frac{p^+}{[k^+-i\d^+][p^+k^-+i\D^-][k^2+i0]} + h.c.
\nn\\
&= \d(1-x) \phi_1^{(\ref{s_virtuals}c)} \,.
\end{align}
It is clear that with the relations in Eq.~(\ref{regul_DeltaDY}), the subtraction of this zero-bin is equivalent to dividing by the soft function in Eq.~(\ref{s_virtuals_c}), proving the equivalence at order $\a_s$ for the virtual contributions.

Let us now consider the real diagrams in fig.~(\ref{n_reals}). The zero-bin of the diagram~(\ref{n_reals}a) is zero.  Diagram~(\ref{n_reals}d) and its zero-bin are zero due to $\bn^2=0$. The zero-bin of diagram~(\ref{n_reals}b) and its Hermitian conjugate~(\ref{n_reals}c) is
\begin{align}
\hat f_{n1,zb}^{(\ref{n_reals}b+\ref{n_reals}c)}&=
-4\pi g^2 C_F \d(1-x)
\int\frac{d^4k}{(2\pi)^4}
\d(k^2)\theta(k^+)\frac{p^+}{[-k^++i\d^+][p^+k^-+i\D^-]}
\d^{(2)}(k_\perp-k_{n\perp})
\nn\\
&+ h.c.
\nn\\
&= \d(1-x) \phi_1^{(\ref{s_reals}b+\ref{s_reals}c)}\,,
\end{align}
which is equivalent to divide by the soft function diagram~(\ref{s_reals}b) and its Hermitian conjugate~(\ref{s_reals}c), given in Eq.~(\ref{eq:jn4b4c}), thanks again to the relation in Eq.~(\ref{regul_DeltaDY}). In conclusion, we have proved that subtracting the zero-bin is equivalent to divide the  naive collinear matrix element by the soft function to first order in $\as$.

%%%%%%%%%%%%%%%%%%%%%%%%%
%%%%%%%%%%%%%%%%%%%%%%%%%
\section{Conclusions}
\label{sec:conc}
%%%%%%%%%%%%%%%%%%%%%%%%%

In this work we have studied the Drell-Yan lepton pair production at  moderately small transverse momentum $q_T$. The analysis was carried out through the framework of the effective field theory via successive a two-step matching procedure: $\textrm{QCD}\to\textrm{SCET-}q_T\to\textrm{SCET-II}$. We established an all-order factorization theorem which allows for a phenomenological study of DY $q_T$ spectrum to be analyzed at energies much larger than $\Lambda_{\rm QCD}$. When considering the double-counting issue of the soft and the naive collinear regions properly, the obtained factorization theorem serves as a guideline towards how the TMDPDF should be defined and what would be its fundamental properties. In our calculations we have used   Wilson lines  defined on-the-light-cone and light-cone singularities appearing in individual Feynman diagrams are regularized with the $\d$-regulator. We have also introduced our result for TMDPDF  using pure DR.
Based on the relations in Eqs.~(\ref{eq:jtmdgeneral}, \ref{eq:jtmd}, \ref{step42}), we were able to define an on-the-light-cone TMDPDF which has the following novel features:
\begin{itemize}
\item It can be integrated over the transverse momentum (of the parton in a parton) to recover the partonic PDF.
\item It is free from rapidity divergences/logarithms.
\item Its evolution is governed by a single parameter  RG evolution equation.
\item It is defined in a gauge invariant way among regular and singular gauges. This definition is obtained from first principles of SCET.
\item It is universal among the DIS and DY kinematics and it can be readily modified to obtain its gluonic version relevant for the proton-proton high-energy collisions.
\end{itemize}

The inclusion of the square root of the soft function in the definition of the TMDPDF has an important consequences:
\begin{itemize}
\item The double counting in the factorization theorem is taken into account.
\item It allows for the separation of UV and IR divergences in the TMDPDF.
\item Even with the subtraction of the square root of the soft function we are able to recover the PDF from TMDPDF.
\end{itemize}

Staying on-the-light-cone has a set of advantages over going off-the-light-cone:
\begin{itemize}
\item The evolution of the TMDPDF is governed with a standard $\m$-RGE.
\item All the results are well-defined without need of using additional parameters.
\item It is consistent with the power counting of SCET. When going off-the-light-cone one needs to introduce a tilted vector $v(\bar {v})$ where there is no specified power counting (in terms of $\lambda$) among its large and small components.
\end{itemize}

The two step factorization is necessary to perform the resummation of logs of $q_T/Q$ and $\Lambda_{QCD}/q_T$ respectively and we have  discussed the resummation procedure in impact parameter. We also commented on the resummed TMDPDF in momentum space.
In the first step of the factorization one gets the usual structure of the cross section given in Eq. ~(\ref{main}). The matrix elements so defined however are not good objects for the second factorization because of the  presence  of mixed  UV/IR   and  rapidity divergences. All these divergences however disappear  in the TMDPDF as we have defined it. The absence of  rapidity divergences allows the resummation of all logs without  the  Collins-Soper like evolution equations.
The second factorization is built up by matching the TMDPDF onto the PDF for large $q_T$.
We have studied the  Wilson coefficients that appear in the second matching which contain, in impact parameters space, $\ln(Q^2/\mu^2)$.
We have shown that  this kind of logs can be exponentiated  in a similar way as was done  in~\cite{Becher:2010tm}, although in our case the concept of collinear anomaly is irrelevant due to the existence of the soft function.

We also considered the TMDPDF with DIS kinematics and pointed out the differences with respect to the DY ones. As mentioned earlier, different Wilson lines are needed for the two kinematical settings. However we established the universality of the TMDPDF in both regimes and argued its validity to all orders in perturbation theory. The gauge invariance of the TMDPDF was also established by computing it  in light-cone gauge with the inclusion of transverse Wilson lines. The fact that all Wilson lines are defined on-the-light-cone only facilitates this computation. We finally comment that our results can be extended to the Higgs boson production at low $q_T$ as well as to spin-dependent non-perturbative hadronic functions, like Sivers and Boer-Mulders ones. The extraction of the resummed TMDPDF from HERA and LHC data, as well as its use to make phenomenological predictions is left for a future consideration.

%%%%%%%%%%%%%%%%%%%%%%%%%%%%%%%%%%%%%%%%
%%%%%%%%%%%%%%%%%%%%%%%%%%%%%%%%%%%%%%%%
\section*{Acknowledgments}
%%%%%%%%%%%%%%%%%%%%%%%%%%%%%%%%%%%%%%%%
This work is supported by the Spanish MEC, FPA2008-00592 and FPA2011-27853-CO2-02.
M.G.E. is supported by the PhD funding program of the
Basque Country Government.
A.I. is supported by BMBF (06RY9191). A.I. has been supported by the
Spanish grant CPAN-ingenio 2010 in the first stage of this work.
I.S. is supported by the Ram\'on y Cajal Program.
M.G.E. would like to thank the Instit\"{u}t f\"{u}r Theoretische Physik, Universit\"at Regensburg, for its support during part of this work was accomplished.
A.I. would like to thank the members of the TQHN group at University of Maryland,
College-Park for their hospitality during which parts of this work were completed.
A.I would like to thank Chul Kim for useful discussions.

\appendix
%%%%%%%%%%%%%%%%%%%%%%%%%%%%%%%%%%%%%%%%
%%%%%%%%%%%%%%%%%%%%%%%%%%%%%%%%%%%%%%%%
\section{Matching the TMDPDF onto PDF with $\d$-Regulator}
\label{sec:app}
%%%%%%%%%%%%%%%%%%%%%%%%%%%%%%%%%%%%%%%%
%%%%%%%%%%%%%%%%%%%%%%%%%%%%%%%%%%%%%%%%

In this appendix we calculate the matching coefficient of the TMDPDF onto the PDF at the intermediate scale using the $\d$-regulator, and show that, as expected, it does not depend on the particular choice of the IR regulator and we get the same result as with pure DR. We also show that this matching coefficient is the same for DY and DIS kinematics, thus establishing its universality to first order in $\as$, since the PDF is universal.

%%%%%%%%%%%%%%%%%%%%%%%%%%%%%%%%%%%%%%%%
%%%%%%%%%%%%%%%%%%%%%%%%%%%%%%%%%%%%%%%%
\subsection{Drell-Yan Kinematics}
\label{sec:app1a}
%%%%%%%%%%%%%%%%%%%%%%%%%%%%%%%%%%%%%%%%
%%%%%%%%%%%%%%%%%%%%%%%%%%%%%%%%%%%%%%%%

The virtual diagrams for the PDF are the same as for the naive collinear matrix element that enters into the definition of the TMDPDF, Fig.~(\ref{n_virtuals}). From Eqs.~(\ref{1a}, \ref{eq:jn1c}) we get
\begin{align}
{\cal Q}_n^{(\ref{n_virtuals}a)}&=
\frac{\alpha_s C_F}{2\pi}
\d(1-x)
\le[ \frac{1}{\veuv}+\ln\frac{\m^2}{\D^-}+\frac{1}{2} \ri] \,,
\end{align}
and
\begin{align}
{\cal Q}_n^{(\ref{n_virtuals}c)}&=
\frac{\a_s C_F}{2\pi}
\d(1-x)
\left[
\frac{2}{\veuv}\ln\frac{\D^+}{Q^2} + \frac{2}{\veuv} - \ln^2\frac{\D^+}{Q^2} -
2\ln\frac{\D^+}{Q^2}\ln\frac{\D^-}{\m^2} - 2\ln\frac{\D^-}{\m^2} + 2 - \frac{7\pi^2}{12}
\right]
\,.
\end{align}
The real diagrams are the same as in Fig.~(\ref{n_reals}), from which we get
\begin{align}
{\cal Q}_n^{(\ref{n_reals}a)}&=
2\pi g^2 C_F p^+ \int\frac{d^dk}{(2\pi)^d}
\d(k^2)\theta(k^+)\frac{2(1-\ve)|\vec k_\perp|^2}{[(p-k)^2+i\D^-][(p-k)^2-i\D^-]}
\d\le((1-x)p^+-k^+\ri)
\nn\\
&=
\frac{\as C_F}{2\pi} (1-x) \left[ \frac{1}{\veuv} + \ln\frac{\m^2}{\D^-} - 1 - \ln(1-x)\right]\,,
\end{align}
and
\begin{align}
{\cal Q}_n^{(\ref{n_reals}b+\ref{n_reals}c)}&=
-4\pi g^2 C_F p^+  \m^{2\ve}\int\frac{d^dk}{(2\pi)^d}
\d(k^2)\theta(k^+)\frac{p^+-k^+}{[k^++i\d^+][(p-k)^2+i\D^-]}
\nn\\
&\times
\d\le((1-x)p^+-k^+\ri) + h.c.
\nn \\
&=
\frac{\as C_F}{2\pi} \left[
\left( \frac{1}{\veuv} + \ln\frac{\m^2}{\D^-} \right)
\left( \frac{2x}{(1-x)_+} - 2\d(1-x)\ln\frac{\D^+}{Q^2} \right)
\right.
\nn\\
&\left.
- 2\d(1-x)\left(1-\frac{\pi^2}{24}-\frac{1}{2}\ln^2\frac{\D^+}{Q^2}\right)
+ \frac{\pi^2}{2} \d(1-x)
\right]\,,
\end{align}
where we have used $\overline {\textrm{MS}}$-scheme, $\m^2\to\m^2e^{\g_E}/(4\pi)$, and the following relations when $\d^+/p^+ \ll 1$,
\begin{align}\label{eq:distributions}
\frac{x}{(1-x)+i\d^+/p^+}+\frac{x}{(1-x)-i\d^+/p^+} &= \frac{2x(1-x)}{(1-x)^2+(\d^+/p^+)^2}
= \frac{2x}{(1-x)_+} - 2\d(1-x)\ln\frac{\d^+}{p^+}\,,
\nn\\
\frac{x(1-x)^{-\ve}}{(1-x) + i\d^+/p^+} + \frac{x(1-x)^{-\ve}}{(1-x) - i\d^+/p^+} &=
2\left[ \frac{x}{(1-x)_+} - \d(1-x)\ln\frac{\d^+}{p^+}
\right.
\nn\\
&\left.
- \ve\d(1-x)\left(1-\frac{\pi^2}{24}-\frac{1}{2}\ln^2\frac{\d^+}{p^+}\right) \right]+{\cal O}(\varepsilon^2)\,,
\nn\\
\frac{x}{(1-x) + i\d^+/p^+} - \frac{x}{(1-x) - i\d^+/p^+} &= -i\pi\d(1-x)
\end{align}
Combining the virtual and real contributions we get the PDF to first order in $\as$
\begin{align}
\label{eq:pdfdelta}
{\cal Q}_n(x;\m) &= \delta(1-x)+\frac{\as C_F}{2\pi} \left[
{\cal P}_{q/q} \left( \frac{1}{\veuv} - \ln\frac{\D^-}{\m^2}\right)
\right.
\nn\\
&\left.
- \frac{1}{4}\d(1-x) - (1-x)\left[1+\ln(1-x)\right]
\right]\,.
\end{align}

The virtual part of the TMDPDF in momentum space was given in Eq.~(\ref{eq:jvn}), and in impact parameter space it reads
\begin{align}
\tilde j_{n1}^v &=
\frac{\alpha_s C_F}{2 \pi}
\d(1-x)
\left[
\frac{1}{\veuv^2} + \frac{1}{\veuv} \left( \frac{3}{2} + \ln\frac{\mu^2\D^+}{Q^2\D^-} \right)
\right.
\nn\\
&\left.
- \frac{3}{2}\ln\frac{\D^-}{\mu^2} - \frac{1}{2}\ln^2\frac{\D^+\D^-}{Q^2\m^2}
+ \ln^2\frac{\D^-}{\m^2} + \frac{7}{4}-\frac{\pi^2}{3}\right]\,.
\end{align}
The Fourier transform of diagrams given in Eqs.~(\ref{eq:jn3a}, \ref{eq:jn3b3c}, \ref{eq:jn4b4c}), while keeping the $\D$'s to regulate the IR divergences, are
\begin{align}
\tilde{\hat f}_{n1}^{(\ref{n_reals}a)}&=
\frac{\as C_F}{2\pi}(1-x)\, \ln\frac{4e^{-2\g_E}}{\D^-(1-x) b^2}\,,
\end{align}

\begin{align}
\tilde{\hat f}_{n1}^{(\ref{n_reals}b+\ref{n_reals}c)}&=
\frac{\as C_F}{2\pi} \left[
\ln\frac{4e^{-2\g_E}}{\D^-b^2}
\left( \frac{2x}{(1-x)_+}-2\d(1-x)\ln\frac{\D^+}{Q^2} \right)
+\frac{\pi^2}{2}\d(1-x)
\right.
\nn\\
&\left.
-2\d(1-x) \left(
1 - \frac{\pi^2}{24} - \frac{1}{2}\ln^2\frac{\D^+}{Q^2}
\right)
\right]\,,
\end{align}
and
\begin{align}
\tilde \phi_1^{(\ref{s_reals}b+\ref{s_reals}c)}&=
\frac{\as C_F}{2\pi}
\left( \ln^2\frac{4e^{-2\g_E}Q^2}{\D^+\D^- b^2} + \frac{2\pi^2}{3} \right)\,,
\end{align}
In the above we have used the following identities in $d=2-2\ve$:
\begin{align}
\int d^d\vec k_\perp e^{i\vec k_\perp\cdot \vec b_\perp}
f(|\vec k_\perp|)
&=
|\vec b_\perp|^{-d} (2\pi)^\frac{d}{2} \int_0^\infty dy\, y^\frac{d}{2} J_{\frac{d}{2}-1}(y)\,
f\left(\frac{y}{|\vec b_\perp|}\right)\,,
\nn\\
\int d^d\vec k_\perp e^{i\vec k_\perp\cdot \vec b_\perp}
f(|\vec k_\perp|)\, \ln|\vec k_\perp|^2
&=
|\vec b_\perp|^{-d} (2\pi)^\frac{d}{2} \int_0^\infty dy\, y^\frac{d}{2} J_{\frac{d}{2}-1}(y)\,
f\left(\frac{y}{|\vec b_\perp|}\right) \ln\frac{y^2}{|\vec b_\perp|^2}\,,
\end{align}
and also
\begin{align}
\int d^d\vec k_\perp e^{i\vec k_\perp\cdot \vec b_\perp}
\frac{1}{|\vec k_{\perp}|^2-i\L^2}
&=
\pi\, \ln\frac{4e^{-2\g_E}}{-i\L^2 b^2}\,,
\nn\\
\int d^d\vec k_\perp e^{i\vec k_\perp\cdot \vec b_\perp}
\frac{|\vec k_{\perp}|^2}{|\vec k_{\perp}|^4+\L^4}
&=
\pi\, \ln\frac{4e^{-2\g_E}}{\L^2 b^2}
\,,
\nn\\
\int d^d\vec k_\perp e^{i\vec k_\perp\cdot \vec b_\perp}
\frac{1}{|\vec k_{\perp}|^2-\L^2} \ln\frac{\L^2}{|\vec k_{\perp}|^2}
&=
\pi \left(
-\frac{1}{2}\ln^2\frac{4e^{-2\g_E}}{\L^2 b^2} - \frac{\pi^2}{3}
\right)\,,
\end{align}
when $\L\to 0$.

Finally, setting $\D^\pm=\D$, the TMDPDF in impact parameter space to first order in $\as$ is
\begin{align}
\tilde j_{n}
&={\cal Q}_n+\frac{\alpha_s C_F}{2\pi}\Big[ -L_T{\cal P}_{q/q}+(1-x) -\delta(1-x)
\left(\frac{1}{2}L_T^2-\frac{3}{2}L_T+L_T\ln\frac{Q^2}{\mu^2}+\frac{\pi^2}{12}\right)\Big]\,,
\end{align}
where ${\cal Q}_n$ is  the PDF given in Eq.~(\ref{eq:pdfdelta}) and the remaining part exactly equals the OPE matching coefficient calculated in pure DR given in Eq.~(\ref{coeff}).

%%%%%%%%%%%%%%%%%%%%%%%%%%%%%%%%%%%%%%%%
%%%%%%%%%%%%%%%%%%%%%%%%%%%%%%%%%%%%%%%%
\subsection{DIS Kinematics}
\label{sec:app1b}
%%%%%%%%%%%%%%%%%%%%%%%%%%%%%%%%%%%%%%%%
%%%%%%%%%%%%%%%%%%%%%%%%%%%%%%%%%%%%%%%%
For DIS kinematics the operator definition of the PDF changes, as we showed in Sec.~\ref{sec:universality},
\begin{align}
{\cal Q}^{DIS}_n(x;\m) = \frac{1}{2} \int \frac{dy^-}{2\pi} e^{-i\frac{1}{2}y^-xp^+}
\left.
\sandwich{p}{\overline {\tilde\chi}_n(0^+,y^-,\vec 0_\perp) \frac{\bnslash}{2}{\tilde\chi}_n^\dagger(0^+,0^-,\vec 0_\perp)}{p}
\ri|_\textrm{zb~ included}\,,
\end{align}
where $\tilde \chi = \tilde W_n^\dagger \xi_n$ and $\tilde W_n^\dagger$ is the collinear Wilson line defined in Sec.~\ref{sec:universality}. In the following we show to first order in $\as$ that the PDF is universal, as expected, although its operator definition changes for DY and DIS kinematics.

\begin{align}
{\cal Q}_n^{DIS(\ref{n_virtuals}a)}&=
\frac{\alpha_s C_F}{2\pi}\d(1-x)
\le[ \frac{1}{\veuv}+\ln\frac{\m^2}{\D^-}+\frac{1}{2} \ri] \,,
\end{align}
and
\begin{align}
{\cal Q}_n^{DIS(\ref{n_virtuals}c)}&=
\frac{\a_s C_F}{2\pi}\d(1-x)
\left[
\frac{2}{\veuv}\ln\frac{\D^+}{Q^2} + \frac{2}{\veuv} - \ln^2\frac{\D^+}{Q^2} -
2\ln\frac{\D^+}{Q^2}\ln\frac{\D^-}{\m^2} - 2\ln\frac{\D^-}{\m^2} + 2 + \frac{5\pi^2}{12}
\right]
\,.
\end{align}
The real diagrams are the same as in Fig.~(\ref{n_reals}), from which we get
\begin{align}
{\cal Q}_n^{DIS(\ref{n_reals}a)}&=
2\pi g^2 C_F p^+ \int\frac{d^dk}{(2\pi)^d}
\d(k^2)\theta(k^+)\frac{2(1-\ve)|k_\perp|^2}{[(p-k)^2+i\D^-][(p-k)^2-i\D^-]}
\d\le((1-x)p^+-k^+\ri)
\nn\\
&=
\frac{\as C_F}{2\pi} (1-x) \left[ \frac{1}{\veuv} + \ln\frac{\m^2}{\D^-} - 1 - \ln(1-x)\right]\,,
\end{align}
and
\begin{align}
{\cal Q}_n^{DIS(\ref{n_reals}b+\ref{n_reals}c)}&=
-4\pi g^2 C_F p^+  \m^{2\ve}\int\frac{d^dk}{(2\pi)^d}
\d(k^2)\theta(k^+)\frac{p^+-k^+}{[k^+-i\d^+][(p-k)^2+i\D^-]}
\nn\\
&\times
\d\le((1-x)p^+-k^+\ri) + h.c.
\nn \\
&=
\frac{\as C_F}{2\pi} \left[
\left( \frac{1}{\veuv} + \ln\frac{\m^2}{\D^-} \right)
\left( \frac{2x}{(1-x)_+} - 2\d(1-x)\ln\frac{\D^+}{Q^2} \right)
\right.
\nn\\
&\left.
- 2\d(1-x)\left(1-\frac{\pi^2}{24}-\frac{1}{2}\ln^2\frac{\D^+}{Q^2}\right)
- \frac{\pi^2}{2} \d(1-x)
\right]\,,
\end{align}
Combining the virtual and real contributions we get the PDF to first order in $\as$
\begin{align}
{\cal Q}^{DIS}_n(x;\m) &={\cal Q}^{DY}_n(x;\m)=
\delta(1-x)+\frac{\as C_F}{2\pi} \left[
{\cal P}_{q/q} \left( \frac{1}{\veuv} - \ln\frac{\D^-}{\m^2}\right)
\right.
\nn\\
&\left.
-\frac{1}{4}\d(1-x) - (1-x)\left[1+\ln(1-x)\right]
\right]\,,
\end{align}
which is the same as in Eq.~(\ref{eq:pdfdelta}) for DY kinematics.

The virtual part of the TMDPDF for DIS kinematics is
\begin{align}
\tilde j_{n1}^{DIS,v} &=
\frac{\alpha_s C_F}{2 \pi}
\d(1-x)
\left[
\frac{1}{\veuv^2} + \frac{1}{\veuv} \left( \frac{3}{2} + \ln\frac{\mu^2\D^+}{Q^2\D^-} \right)
\right.
\nn\\
&\left.
- \frac{3}{2}\ln\frac{\D^-}{\mu^2} - \frac{1}{2}\ln^2\frac{\D^+\D^-}{Q^2\m^2}
+ \ln^2\frac{\D^-}{\m^2} + \frac{7}{4}+\frac{\pi^2}{6}\right]\,.
\end{align}
The real diagrams and their Fourier transforms are
\begin{align}
\hat f_{n1}^{DIS(\ref{n_reals}a)}&=
2\pi g^2 C_F p^+ \int\frac{d^dk}{(2\pi)^d}
\d(k^2)\theta(k^+)\frac{2(1-\ve)|\vec k_\perp|^2}{[(p-k)^2+i\D^-][(p-k)^2-i\D^-]}
\nn\\
&\times
\d\le((1-x)p^+-k^+\ri) \d^{(2)}(\vec k_\perp+\vec k_{n\perp})
\nn\\
&=
\frac{2\a_s C_F}{(2\pi)^{2-2\ve}}
(1-\ve)(1-x)
\frac{|\vec k_{n\perp}|^2}{\le| |\vec k_{n\perp}|^2-i\D^-(1-x)\ri|^2}\,,
\end{align}

\begin{align}
\tilde{\hat f}_{n1}^{DIS(\ref{n_reals}a)}&=
\frac{\as C_F}{2\pi}(1-x)\, \ln\frac{4e^{-2\g_E}}{\D^-(1-x) b^2}\,,
\end{align}

\begin{align}
\hat f_{n1}^{DIS(\ref{n_reals}b+\ref{n_reals}c)}&=
-4\pi g^2 C_F p^+  \int\frac{d^dk}{(2\pi)^d}
\d(k^2)\theta(k^+)\frac{p^+-k^+}{[k^+-i\d^+][(p-k)^2+i\D^-]}
\nn\\
&\times
\d\left((1-x)p^+-k^+\right) \d^{(2)}(\vec k_\perp+\vec k_{n\perp}) + h.c.
\nn\\
&=
\frac{2\alpha_s C_F}{(2\pi)^{2-2\ve}}
\left[\frac{x}{(1-x)-i\d^+/p^+}\right]
\left[
\frac{1}{|\vec k_{n\perp}|^2-i\D^-(1-x)}
\right]
+ h.c.\,,
\end{align}

\begin{align}
\tilde{\hat f}_{n1}^{DIS(\ref{n_reals}b+\ref{n_reals}c)}&=
\frac{\as C_F}{2\pi} \left[
\ln\frac{4e^{-2\g_E}}{\D^-b^2}
\left( \frac{2x}{(1-x)_+}-2\d(1-x)\ln\frac{\D^+}{Q^2} \right)
-\frac{\pi^2}{2}\d(1-x)
\right.
\nn\\
&\left.
-2\d(1-x) \left(
1 - \frac{\pi^2}{24} - \frac{1}{2}\ln^2\frac{\D^+}{Q^2}
\right)
\right]\,,
\end{align}

\begin{align}
\phi_1^{DIS(\ref{s_reals}b+\ref{s_reals}c)}&=
-4\pi g^2 C_F  \int\frac{d^dk}{(2\pi)^d}
\d^{(2)}(\vec k_\perp+\vec k_{n\perp}) \d(k^2)\theta(k^+) \frac{1}{[k^+-i\d^+][-k^-+i\d^-]} + h.c.
\nn\\
&=
-\frac{4\a_s C_F}{(2\pi)^{2-2\ve}}
\frac{1}{|\vec k_{n\perp}|^2+\d^+\d^-} \ln\frac{\d^+\d^-}{|\vec k_{n\perp}|^2}\,,
\end{align}

\begin{align}
\tilde \phi_1^{DIS(\ref{s_reals}b)}&=
 \frac{\as C_F}{2\pi}
\left( \ln^2\frac{4e^{-2\g_E}Q^2}{\D^+\D^- b^2} - \frac{\pi^2}{3} \right)\,,
\end{align}
In the above we have used the following identity:
\begin{align}
\int d^d\vec k_\perp e^{i\vec k_\perp\cdot \vec b}
\frac{1}{|\vec k_{\perp}|^2+\L^2} \ln\frac{\L^2}{|\vec k_{\perp}|^2}
&=
\pi \left(
-\frac{1}{2}\ln^2\frac{4e^{-2\g_E}}{\L^2 b^2} + \frac{\pi^2}{6}
\right)\,,
\end{align}
when $\L\to 0$.

Finally, setting $\D^\pm=\D$, the total TMDPDF in impact parameter space to first order in $\as$ is
\begin{align}
\tilde j_{n}^{DIS}
&={\cal Q}^{DIS}_n+\frac{\alpha_s C_F}{2\pi}\Big[ -L_T{\cal P}_{q/q}+(1-x) -\delta(1-x)
\left(\frac{1}{2}L_T^2-\frac{3}{2}L_T+L_T\ln\frac{Q^2}{\mu^2}+\frac{\pi^2}{12}\right)\Big]\,,
\end{align}
where ${\cal Q}_n$ is the PDF given in Eq.~(\ref{eq:pdfdelta}) and the remaining part exactly equals the OPE matching coefficient calculated in DY kinematics.

%%%%%%%%%%%%%%%%%%%%%%%%%%%%%%%%%%%%%%%%
%%%%%%%%%%%%%%%%%%%%%%%%%%%%%%%%%%%%%%%%
\section{Quark Form Factor in Full QCD with $\d$-Regulator}
\label{sec:app2}
%%%%%%%%%%%%%%%%%%%%%%%%%%%%%%%%%%%%%%%%
%%%%%%%%%%%%%%%%%%%%%%%%%%%%%%%%%%%%%%%%
In this appendix we calculate the Quark Form Factor (QFF) in full QCD with the $\d$-regulator for DIS and DY kinematics. We will show that the real part of the QFF is the same in both regimes, and that the imaginary part changes sign. We set $\D^\pm=\D$.

For DIS setting the vertex correction is
\begin{align}
V^{\a,DIS} &=-ig^2 C_F \m^{2\e} \int \frac{d^dk}{(2\pi)^d}
\frac{\bar u_{n} \g^\m (\pslash-\kslash) \g^\a (\bpslash-\kslash) \g_\m u_{\bn}}
{[(p-k)^2+i\D] [(\bp-k)^2+i\D] [k^2+i0]}
\nn\\
&=
\frac{\as C_F}{4\pi} \g^\a_\perp \left(
\frac{1}{\ve} + 2i\pi\ln\frac{\D}{Q^2} - 2\ln^2\frac{\D}{Q^2}
- 4\ln\frac{-i\D}{\m^2} + 3\ln\frac{Q^2}{\m^2} - 4 + \frac{\pi^2}{2}
\right)\,,
\end{align}
where we have taken the large $Q^2$ limit and only the terms proportional to $\g^\a_\perp$.
The WFR is the following,
\begin{align}
I_w &=
-g^2 C_F\m^{2\e} \int
\frac{d^dk}{(2\pi)^d} \frac{\g^\m(\pslash-\kslash)\g_\m}{[(p-k)^2+i\D][k^2+i0]} =
i\pslash \frac{\as C_F}{4\pi} \left( \frac{1}{\ve} + \frac{1}{2} + \ln\frac{\m^2}{-i\D} \right)\,.
\end{align}
The final result for the QFF calculated in QCD, with the $\d$-regulator and for DIS kinematics is
\begin{align}
\label{eq:currentdis}
<p_{n} | J^\a | \bp_{\bn}> ^{DIS} &=
V^{\a,DIS} -\frac{1}{2} \frac{I_w(\bp)}{i\bpslash}\g^\a_\perp - \frac{1}{2} \frac{I_w(p)}{i\pslash}\g^\a_\perp =
\nn\\
&=
\g^\a_\perp \left[ 1+\frac{\a C_F}{4\pi} \left(
- 2\ln^2\frac{\D}{Q^2} - 3\ln\frac{-i\D}{Q^2} + 2i\pi\ln\frac{\D}{Q^2}
- \frac{9}{2} + \frac{\pi^2}{2}
 \right)\right]\,.
\end{align}

The analogous results for DY setting are the following,
\begin{align}
V^{\a,DY}&=+ig^2 C_F \m^{2\e} \int \frac{d^dk}{(2\pi)^d}
\frac{\bar u_{n} \g^\m (\pslash-\kslash) \g^\a (\bpslash+\kslash) \g_\m u_{\bn}}
{[(p-k)^2+i\D^-] [(\bp+k)^2+i\D^+] [k^2+i0]}
\nn\\
&=
\frac{\as C_F}{4\pi} \g^\a_\perp \left(
\frac{1}{\ve} - 2i\pi\ln\frac{\D}{Q^2} - 2\ln^2\frac{\D}{Q^2}
- 4\ln\frac{-i\D}{\m^2} + 3\ln\frac{-Q^2}{\m^2} - 4 + \frac{\pi^2}{2}
\right)\,,
\end{align}
and the same WFR as in DIS. The final result for the QFF is then
\begin{align}
\label{eq:currentdy}
<p_{n} | J^\a | \bp_{\bn}>^{DY} &=
V^{\a,DY} -\frac{1}{2} \frac{I_w(\bp)}{i\bpslash}\g^\a_\perp - \frac{1}{2} \frac{I_w(p)}{i\pslash}\g^\a_\perp =
\nn\\
&=
\g^\a_\perp \left[ 1+\frac{\a C_F}{4\pi} \left(
- 2\ln^2\frac{\D}{Q^2} - 3\ln\frac{-i\D}{-Q^2} - 2i\pi\ln\frac{\D}{Q^2}
- \frac{9}{2} + \frac{\pi^2}{2}
 \right)\right]\,.
\end{align}

In conclusion, the real part of the current calculated with the $\d$-regulator is the same for DY and DIS kinematics. This is due to the fact that our regulator, $i\D$, is imaginary, so changing from space-like (DIS) to time-like (DY) does not change the real part.

Notice that Eq.~(\ref{eq:currentdy}) is consistent with Eq.~(\ref{eq:mqcdv}) when adding its Hermitian conjugate. Combining Eq.~(\ref{eq:currentdis}) with its Hermitian conjugate and the virtual contributions $j_n^{v,DIS}$ and $j_\bn^{v,DIS}$ in Eq.~(\ref{eq:resultsdis}) we recover the hard matching coefficient for DIS between QCD and SCET (Eq.~(20) in \cite{Manohar:2003vb} plus its Hermitian conjugate).

%%%%%%%%%%%%%%%%%%%%%%%%%%%%%%%%%%%%%%%%
%%%%%%%%%%%%%%%%%%%%%%%%%%%%%%%%%%%%%%%%
%%%%%%%%%%%%%%%%%%%%%%%%%%%%%%%%%%%%%%%%

%%%%%%%%%%%%%%%%%%%%%%%%%%%%%%%%%
%%%%%%%%%%%%%%%%%%%%%%%%%%%%%%%%%
%%%%%%%%%%%%%%%%%%%%%%%%%%%%%%%%%
\end{document}